\documentclass[5p,twocolumn,authoryear]{elsarticle}
\usepackage{amssymb,amsmath}
\usepackage{mathrsfs}
\usepackage{caption,subcaption}
\usepackage{graphicx}
\usepackage{color,xcolor}
\usepackage{epstopdf}
\usepackage{enumitem}
\usepackage{bbm}

\usepackage{hyperref}
\hypersetup{
    colorlinks = true,
    allbordercolors = {white},
    allcolors = {blue},
}

\newtheorem{corollary}{Corollary}
\newtheorem{remark}{Remark}
\newproof{proof}{Proof}

\newtheorem{theorem}{Theorem}
\newtheorem{lemma}{Lemma}

\newtheorem{definition}{Definition}

\newtheorem{assum}{Assumption}
\newtheorem{proposition}{Proposition}

\def\epf{\blacksquare}

\def\g{\mathcal{G}}
\def\v{\mathcal{V}}
\def\e{\mathcal{E}}
\def\w{\mathscr{\bf w}}

\def\r{\mathbb R}
\def\ones{\mathbbm{1}}

\def\la{\lambda}

\def\ve{\varepsilon}

\def\M{\mathfrak{M}}

\def\be{\beta}

\usepackage{amsopn}
\DeclareMathOperator{\diag}{diag}
\DeclareMathOperator{\sgn}{sgn}

\def\epf{$\blacksquare$}

\usepackage{tikz}
\usepackage{tkz-graph}
\usetikzlibrary{shapes}
\GraphInit[vstyle = Shade]
\tikzset{
  LabelStyle/.style = { rectangle, rounded corners, draw,
                        minimum width = 2em, fill = yellow!50,
                        text = red, font = \bfseries },
  VertexStyle/.append style = { minimum width = 1em,
                                inner sep=2pt,
                                font = \large\bfseries},
  EdgeStyle/.append style = {->, bend left} }

\def\be{\begin{equation}}
\def\ee{\end{equation}}
\def\ben{\begin{equation*}}
\def\een{\end{equation*}}
\newcommand{\dfb}{\stackrel{\Delta}{=}}

\def\r{\mathbb R}
\def\w{\mathscr{\bf w}}
\def\ones{\mathbbm{1}}

\journal{Annual Reviews in Control}
\begin{document}

\begin{frontmatter}
\title{Recurrent Averaging Inequalities \\in Multi-Agent Control and Social Dynamics Modeling}

\author[First,Second]{Anton V.~Proskurnikov\corref{cor2}\fnref{fn2}}
\ead{anton.p.1982@ieee.org}

\author[First]{Giuseppe C. Calafiore}
\ead{giuseppe.calafiore@polito.it}

\author[Third]{Ming Cao}
\ead{m.cao@rug.nl}

\cortext[cor2]{Corresponding author.}

\fntext[fn2]{Some results of the paper were reported~\citep{ProCao2017-3} on IEEE Conference on
Decision and Control CDC 2017, Melbourne, Australia.}

\address[First]{Department of Electronics and Telecommunications,
Politecnico di Torino, Turin, Italy}
\address[Second]{Institute for Problems of Mechanical Engineering of the Russian Academy of Sciences (IPME RAS), St. Petersburg, Russia}
\address[Third]{Faculty of Science and Engineering, University of Groningen, Groningen, The Netherlands}

\begin{abstract}
Many multi-agent control algorithms and dynamic agent-based models arising in natural and social sciences are based on the principle of \emph{iterative averaging}. Each agent is associated to a value of interest, which may represent, for instance, the opinion of an individual in a social group, the velocity vector of a mobile robot in a flock, or the measurement of a sensor within a sensor network. This value is updated, at each iteration, to a weighted average
of itself and of the values of the adjacent agents. It is well known that, under natural assumptions on the network's graph connectivity, this local averaging procedure eventually leads to global consensus, or synchronization of the values at all nodes. Applications of iterative averaging include, but are not limited to, algorithms for distributed optimization, for solution of linear and nonlinear equations, for multi-robot coordination and for opinion formation in social groups.
Although these algorithms have similar structures, the mathematical techniques used for their analysis are diverse, and conditions for their convergence and differ from case to case. In this  paper, we review
many of these algorithms and we
show that their  properties  can be analyzed in a unified way by using a novel tool based on recurrent averaging inequalities (RAIs). We develop a theory of RAIs and apply it to the analysis of several important
multi-agent algorithms recently proposed in the literature.
\end{abstract}

\begin{keyword}
 Distributed algorithm, opinion dynamics, multi-agent control, consensus
\end{keyword}

\end{frontmatter}

\section{Introduction}

In last two decades problems of distributed control and coordination in multi-agent systems (MAS)~\citep{MesbahiEgerBook,ShammaBook,RenBeardBook,RenCaoBook,LewisBook} have attracted an enormous attention from the research community. The fundamental problem in MAS theory is how a complex cooperative goal (e.g., coordinated motion such as flocking or swarming, uniform deployment over some area, collective decision making) can be reached by a team of relatively simple autonomous agents. Usually, agents have restricted computational capabilities and limited information about the team as a whole; each agent can interact only with a few neighboring agents. The neighborship relations are conveniently represented by the interaction graph, or ``topology'' of the multi-agent system. Agent-based models proved to be useful in the study of complex phenomena arising in nature and society, see, e.g.,~\citep{LEONARD2014171,ProTempo:2017-1,BulloBook-Online}.

\begin{figure}[tb]
\centering
\includegraphics[width=0.35\columnwidth]{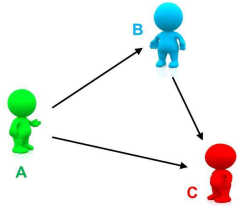}
\caption{French's model of opinion formation in a group of $n=3$ actors. $A$'s opinion is displayed to $B$ and $C$, $B$'s opinion is displayed to $C$ but is hidden from $A$, $C$'s opinion is displayed to nobody. The corresponding dynamical system obeys the equations $x_A(k+1)=x_A(k)$, $x_B(k+1)=\frac{x_A(k)+x_B(k)}{2}$, $x_C(k+1)=\frac{x_A(k)+x_B(k)+x_C(k)}{3}$.}\label{fig.french}
\end{figure}

Many multi-agent algorithms considered in the literature are based, explicitly or implicitly, on the idea of \emph{iterative averaging}. Discrete-time dynamics of iterative averaging date back to the seminal model of opinion diffusion in social groups introduced by~\citep{French:1956} and thoroughly studied in~\citep{Harary:1959,HararyBook:1965}. In French's model, each agent (social actor) corresponds to a node of a graph and keeps an \emph{opinion} -- a real number standing for some quantity of interest or cognitive orientation towards some object~\citep{Friedkin:2015}. At each step of the opinion iteration, the actors simultaneously display their opinions to the neighbors in the graph and update their opinions based on the information displayed to them: a new value of an actor's opinion is computed as the mean of its previous value and the opinions displayed by the neighbors (Fig.~\ref{fig.french}).
French's model can be written as
\be\label{eq.degroot_scal}
x_i(k+1)=\sum_{j=1}^nw_{ij}x_j(k),\; i=1,\ldots,n; \; k=0,1,\ldots
\ee
where
$x_i(k)$ denotes the opinion of actor $i$ at time $k$, and $w_{ij}$ may be interpreted as the relative strength of influence of agent $j$ on agent $i$ (the diagonal entries measure self-influence of the agents).  The weights  $w_{ij}$ are nonnegative 
and $\sum_{j=1}^nw_{ij}=1$ for all  $i$. As shown in~\citep{Harary:1959,HararyBook:1965}, the iteration of process~\eqref{eq.degroot_scal} typically  establishes eventual consensus (unanimity) of opinions.

Collecting the weights in a (row-stochastic) matrix $W=(w_{ij})$, and defining
the vector $x(k)\in\r^{n}$ of opinions at time $k$, we rewrite model \eqref{eq.degroot_scal} in vector format as
\be\label{eq.degroot}
x(k+1)=Wx(k),\quad k=0,1,\ldots
\ee
Model~\eqref{eq.degroot}, with a general stochastic matrix $W$, was later independently proposed by~\cite{DeGroot} and~\cite{Lehrer:1976} as a method for reaching rational agreement in expert communities. Consensus in~\eqref{eq.degroot} is equivalent  to the ``full regularity''~\citep{GantmacherVol2}, or SIA (stochastic, indecomposable, aperiodic) property~\citep{Wolfowitz:1963} of the matrix $W$ or, equivalently, regularity of the Markov chain generated by this matrix~\citep{KemenySnellMarkovBook}. In particular, the iterations in~\eqref{eq.degroot} always reach consensus if $W$ is aperiodic and irreducible (or primitive)~\citep{DeGroot,GantmacherVol2}. In graph-theoretic language,
the necessary and sufficient criterion of consensus is the existence of an agent that communicates, directly or indirectly, with all other agents, that is, the graph has a directed spanning tree~\citep{RenBeardBook,BulloBook-Online}.

The behavior of iterative averaging algorithms such as~\eqref{eq.degroot}, however, is more complicated in the case of  \emph{dynamic} interaction graphs, where the matrix of influence weights $W=W(k)$ varies with time
\be\label{eq.degroot-classic}
x(k+1)=W(k)x(k),\quad k=0,1,\ldots
\ee
While such models were initially studied in relation to non-stationary Markov processes~\citep{Dobrushin:56,Hajnal:58,Wolfowitz:1963,Seneta},
they more recently experienced a resurgence
of interest
 driven by agent-based models of synchronous collective motion based on the ``nearest neighbor'' rules~\citep{Reynolds,Vicsek,Saber2006,Morse12}.
 In such models, the interactions between the agents are range-restricted, and the set of an agent's neighbors evolves as the agents move in the space. Consensus in the time-varying algorithm~\eqref{eq.degroot-classic} is equivalent to the ergodicity\footnote{The backward products of stochastic matrices $W(k)\ldots W(0)$ are \emph{ergodic} if they converge, as $k\to\infty$, to a stochastic rank-one matrix.} of the backward infinite matrix products $W(k)\ldots W(0)$, see~\cite{Seneta}. Despite the interest and efforts devoted to the problem of matrix products ergodicity in the literature on probability theory and matrix analysis~\citep{Seneta,LEIZAROWITZ1992189}, a complete solution to this problem remains elusive, and a gap between necessary and sufficient ergodicity conditions still exists~\citep{TOURI20121477,Touri:14,Bolouki:2016}.

In recent years, the iterative averaging model has attracted significant attention from the research community
as a simple algorithm for multi-agent coordination~\citep{Jad:03,Blondel:05,Moro:05,Murray:07,MesbahiEgerBook,RenBeardBook,RenCaoBook,CaoMorse:08}.
Besides the important problem of multi-agent consensus, procedures of iterative averaging lie at the core of many distributed numerical algorithms~\citep{Tsitsiklis} such as, e.g., techniques of distributed estimation and filtering~\citep{Borkar:82, CalAbr:09, Garin2010,BulloBook-Online}, deterministic and stochastic optimization~\citep{Tsitsiklis:86,Nedic:10,LinRen:14,YANG2019_ARC}, load balancing~\citep{Amelina:15}, and solving systems of linear equations~\citep{LiuMorseNedicBasar:14,MouLiuMorse:15,YouSongTempo:16,Wang2019_ARC}. Many models of opinion formation~\citep{ProTempo:2017-1,ProTempo:2018} in social groups stem from the classical model~\eqref{eq.degroot}.

While the mentioned algorithms and dynamical models have similar structures, their properties and the mathematical techniques used for their analysis usually differ.
For instance, under some natural assumptions, the standard condition for consensus in~\eqref{eq.degroot} is the existence of a directed spanning tree in the communication graph~\citep{RenCaoBook,CaoMorse:08}, known also as the quasi-strong connectivity.
At the same time, similar algorithms for \emph{constrained} consensus~\citep{Nedic:10} and linear equations solving~\citep{MouLiuMorse:15,YouSongTempo:16} that are also based on iterative averaging, require strongly connected graphs. Their convergence properties cannot be directly derived from properties of the standard consensus algorithm, and require additional tools. To the best of the authors' knowledge, there is no unified mathematical theory for
averaging-based multi-agent algorithms.

\subsection{Systems of recurrent averaging inequalities}

In this paper, while
reviewing many of the aforementioned averaging-based algorithms, we  demonstrate that they can be analyzed in a unified way by means of
a novel tool that we call a \emph{recurrent averaging inequality} (RAI), which
is defined as an inequality relaxation of the (possibly time-varying) iterations in~\eqref{eq.degroot}, namely,
\be\label{eq.ineq}
x(k+1)\le W(k)x(k), \quad k=0,1,\ldots,  
\ee
where matrices $W(k)$ are row-stochastic.
At a first glance, the system of inequalities~\eqref{eq.ineq} is too ``loose'' to entail any interesting properties of the sequence $\{x(k)\}$.
However, under rather modest conditions of connectivity (for instance, if every matrix $W(k)+\ldots+W(k+T-1)$, where $k\ge 0$ and $T$ is a fixed period, corresponds to a strongly connected graph) the inequality~\eqref{eq.ineq} implies asymptotic consensus of the opinions $x_i(k)$ (which, however, can be achieved at $-\infty$). In many situations where consensus is not established,~\eqref{eq.ineq} provides convergence of the sequence $\{x(k)\}$. Similar properties have been recently obtained~\citep{ProCao:2017} for inequalities 
\be\label{eq.ineq-cont}
\dot x(t)\le -L(t)x(t),
\ee
where $L(t)$ stands for a time-varying Laplacian matrix. The theory developed in~\cite{ProCao:2017} is not directly applicable to discrete-time inequalities~\eqref{eq.ineq}. 

\subsection{Contribution and paper organization}
In this paper we  develop a mathematical theory for the RAI~\eqref{eq.ineq}. In particular, we establish convergence and consensus criteria for any feasible sequence $\{x(k)\}$. Further, we review
a number of multi-agent algorithms and opinion formation models, and
show that they  can be examined in a unified way using the RAI theory, which allows, in particular, to examine Hegselmann-Krause model with informed (``truth-seeking'') agents~\cite{HegselmannKrause:2006}, and to generalize the recent fundamental results~\cite{FullmerMorse2018} on distributed algorithms that compute a common fixed point for a family of paracontractions, thus solving a special system of nonlinear equations.

The paper is organized as follows: Section~\ref{sec.prelim}  introduces the notation and some concepts from graph theory. Section~\ref{sec.classic} recapitulates some classical results concerned with consensus in iterative averaging procedures. Section~\ref{sec.main} presents new results, establishing convergence of the solutions of  the RAI~\eqref{eq.ineq}. In Section~\ref{sec.appl}, we illustrate the results showing applications to  models of opinion formation and distributed algorithms. Section~\ref{sec.proofs} contains the proofs of the main results. Section~\ref{sec.conclus} concludes the paper.

\section{Preliminaries and notation}\label{sec.prelim}
For  positive integers $m,n$, with $n\ge m$, we let $[m:n]\dfb\{m,m+1,\ldots,n\}$. The cardinality of a finite set $I$ is denoted by $|I|$. We use $\ones_n=(1,1,\ldots,1)^{\top}\in\r^n$ to denote a (column) vector of all ones, and $\mathbf{e}_1=(1,0,\ldots,0)^{\top},\ldots,\mathbf{e}_n=(0,0,\ldots,1)^{\top}$ to denote the basis of the unit coordinate vectors.
For two vectors $x,y\in\r^n$ we write
$x\le y$ if $x_i\le y_i\,\forall\, i$ and $x\le 0$ if $x_i\le 0\,\forall\, i$; the reverse inequalities $\geq $ are defined analogously. We define the sign of a real number as follows
\[
\sgn( t ) =
\begin{cases}
1,& t>0\\
-1,& t<0\\
0,& t=0.
\end{cases}
\]
The spectral radius of a matrix $A$ is denoted by $\rho(A)$.

A graph is a pair $\g=(\v,\e)$, where $\v$ is a finite set referred to as the set of \emph{nodes} and $\e\subseteq \v\times\v$ is a set of \emph{arcs}. The arc $(i,j)$ is also denoted by $i\xrightarrow{}j$. We call a graph \emph{undirected} if $\e$ is symmetric in the sense that $(i,j)\in\e\Leftrightarrow (j,i)\in\e$, otherwise the graph is \emph{directed}. A sequence of arcs $i_0\xrightarrow{}i_1\xrightarrow{}i_2\xrightarrow{}\ldots\xrightarrow{}i_s$ is called a \emph{walk} connecting $i_0$ to $i_s$, the number of arcs $s$ is the walk's \emph{length}. A walk that starts and ends at the same node $i_0=i_s$ is said to be a \emph{cycle.} A graph is said to be \emph{periodic} if an integer $p>1$ exists that divides the length of any cycle; otherwise the graph is \emph{aperiodic}. A graph is \emph{strongly connected} if every two distinct nodes are connected by a walk. A graph is \emph{quasi-strongly connected}, or has a directed (out-branching) \emph{spanning tree}~\citep{RenBeardBook} if some node (``root'') is connected by walks to all other nodes. For undirected graphs, strong and quasi-strong connectivity are equivalent (such a graph is said to be \emph{connected}).

\begin{figure}[h]
\begin{subfigure}{0.49\columnwidth}
\centering
\includegraphics[width=\textwidth]{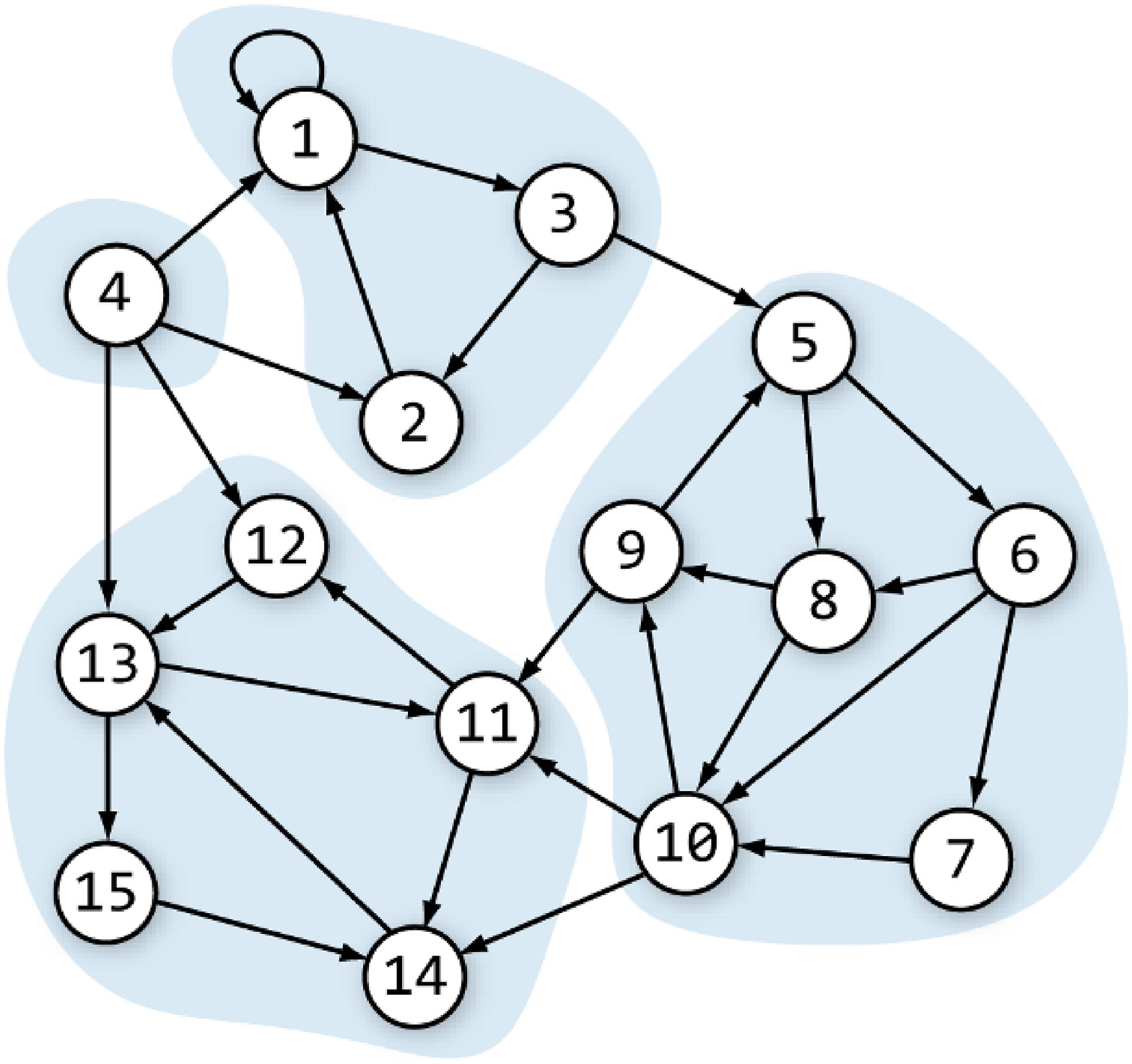}
\caption{}
\end{subfigure}
\begin{subfigure}{0.49\columnwidth}
\centering
\includegraphics[width=\textwidth]{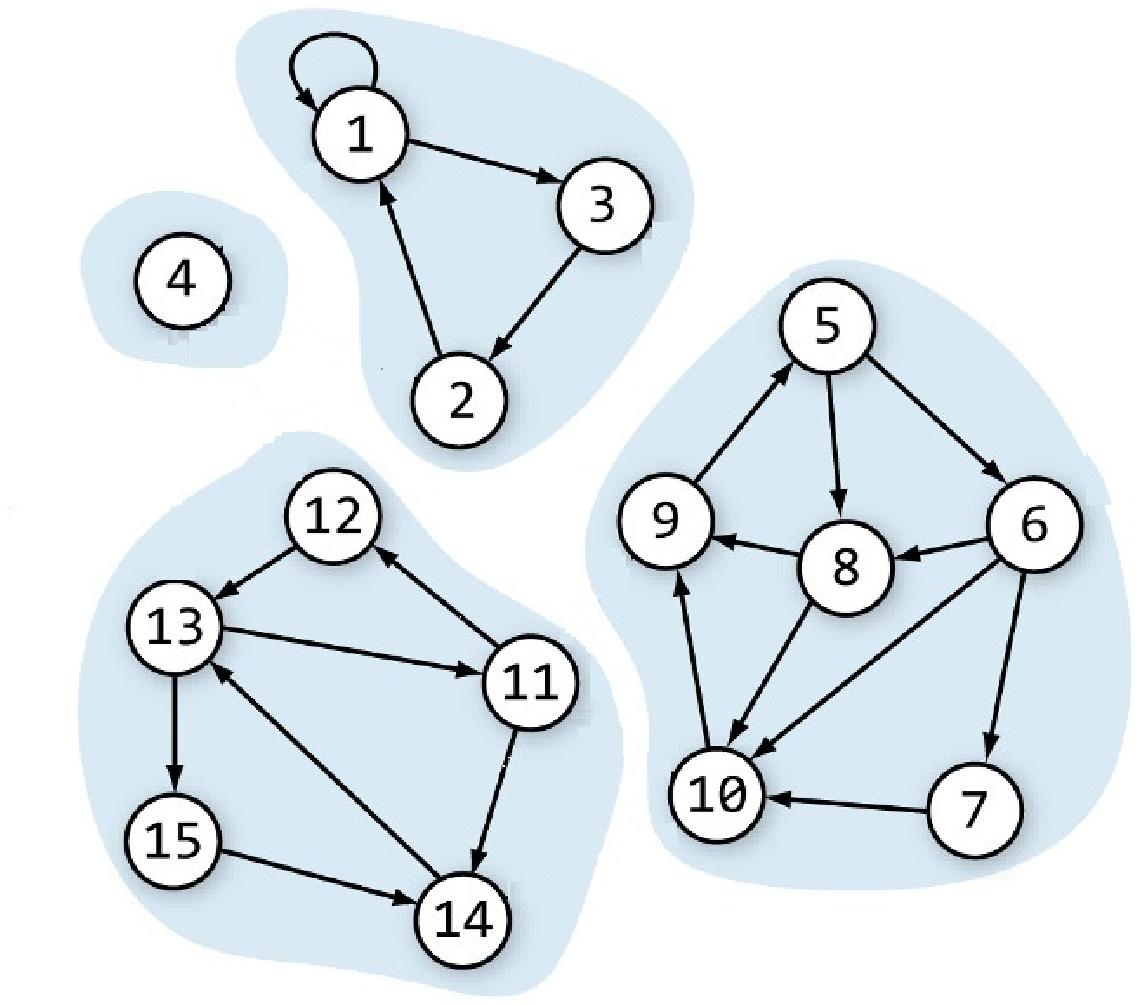}
\caption{}
\end{subfigure}
\caption{Strong components of a directed graph: (a) non-isolated; (b) isolated. In (a), $\{4\}$ is a single source component, $\{11,\ldots,15\}$ is a single sink component.}
\label{fig.strong}
\end{figure}

A graph $(\v',\e')$, where $\v'\subseteq\v$ and $\e'\subseteq\e$ is referred to as a \emph{subgraph} of the graph $(\v,\e)$. A subgraph is said to be a \emph{strongly connected} (or simply \emph{strong}) \emph{component} if it is strongly connected and maximal in the sense that no other node or arc can be added to it without destroying the subgraph's strong connectivity. Each node of $\g$ belongs to at most one strong component. If $\g$ is strongly connected, then it has only one strong component ($\g$ itself). Otherwise, it contains multiple strong components, among which at least one component is a \emph{source component} (no arcs enter it) and at least one component is a \emph{sink component} (no arcs leave it), see Fig.~\ref{fig.strong}. A graph is quasi-strongly connected if and only if it has a single source component.
A strong component can be \emph{isolated}, when it has neither incoming nor outcoming arcs, and thus it is both a source and a sink.
Strong components of undirected graphs are always isolated.

 An arbitrary matrix $A=(a_{ij})\in\r^{n\times n}$
 can be associated with a \emph{signed}
 weighted graph, being the triple $\g[A]=([1:n],E[A],A)$, where $[1:n]$ is the set of nodes, $E[A]\dfb\{(j,i):a_{ij}\ne 0\}$
 is the set of arcs and $a_{ij}$ stands for the (signed) weight or value of arc\footnote{In the models considered below, entry $a_{ij}$ usually quantifies the strength of influence agent $j$ exerts on agent $i$. Following the tradition of multi-agent systems theory~\citep{RenBeardBook,RenCaoBook} and the original work by~\cite{French:1956},
such an influence is depicted by an arc $j\xrightarrow{}i$ rather than $i\xrightarrow{}j$.} $(j,i)$. In this paper, we mainly deal with graphs generated by non-negative matrices: in such a situation, the entry of a matrix is considered as a weight (or value) of the corresponding arc. For nonnegative matrices $B_1,B_2\in\r^{n\times n}$, one has $\e[B_1+B_2]=\e[B_1]\cup\e[B_2]$; the resulting graph $\g[B_1+B_2]$ is called the \emph{union} of the graphs $\g[B_1]$ and $\g[B_2]$. A non-negative matrix $B$ is \emph{irreducible} if $\g[B]$ is strongly connected and \emph{aperiodic} if $\g[B]$ is aperiodic.

The subdivision of nodes into two non-empty disjoint sets $I$ and $J=I^c\dfb[1:n]\setminus I$ is referred to as a \emph{cut} in $\g[B]$. The graph $\g[B]$, corresponding to a non-negative matrix $B$, is said to be \emph{weight-balanced} if the total weights of incoming and outcoming arcs are same for all nodes $\sum_{j}b_{ij}=\sum_{j}b_{ji}\,\forall i$. For any cut $(I,J)$ in such a graph, the balance condition holds as follows
\[
\sum_{i\in I,j\in J}b_{ij}=\sum_{i\in I,j\in J}b_{ji}.
\]
Following~\citep{TsiTsi:13}, we call a graph \emph{cut-balanced} if a constant $C\ge 1$ exists such that
\be\label{eq.cut-balance-static1}
C^{-1}\sum_{i\in I,j\in J}b_{ji}\leq\sum_{i\in I,j\in J}b_{ij}\leq C\sum_{i\in I,j\in J}b_{ji}.
\ee
The following  result characterizes a cut-balanced graph.
\begin{lemma}\label{lem.isolated}
For a nonnegative matrix $B$, the following statements are equivalent:
\begin{enumerate}
\item graph $\g[B]$ is cut-balanced;
\item flows from $I$ to $J$ and from $J$ to $I$ are either both positive or both zero:
\be\label{eq.cut-balance-static}
\sum_{i\in I,j\in J}b_{ij}>0\Longleftrightarrow \sum_{i\in I,j\in J}b_{ji}>0;
\ee
\item all strongly connected components of $\g[B]$ are isolated;
\item  for any $i,j\in[1:n]$, a walk from $i$ to $j$ exists if and only if a walk from $j$ to $i$ exists.
\end{enumerate}
\end{lemma}

\begin{proof}
The implications 1$\Longrightarrow$3 and $3\Longleftrightarrow$4 follow from a more general result~\cite[Lemma~1]{TsiTsi:13}. To prove $4\Longrightarrow$2, note that the left-hand inequality in~\eqref{eq.cut-balance-static} holds if and only if an arc $(j,i)$ from $j\in J$ to $i\in I$ exists.
The walk from $i$ to $j$ (existing due to 4) contains an arc connecting a node from $I$ to a node from $J$, therefore, the right-hand side inequality in~\eqref{eq.cut-balance-static} is also valid.
The right-hand side inequality in~\eqref{eq.cut-balance-static} entails the left-hand side one for the same reason. To prove that 2$\Longrightarrow$1, it suffices to notice that~\eqref{eq.cut-balance-static1} holds with
the constant
\[
C\dfb\max_{(I,J)}\frac{\sum_{i\in I,j\in J}b_{ij}}{\sum_{i\in I,j\in J}b_{ji}},
\]
where the maximum is taken over all possible cuts such that the ratio is well-defined.
\end{proof}


\section{Convergence of classical averaging algorithms}\label{sec.classic}

We start by reviewing some  basic results on stability of the iterative averaging algorithm~\eqref{eq.degroot-classic},
where ${W(k)}_{k\ge 0}$ is a sequence of $n\times n$ row-stochastic matrices, corresponding to a dynamic communication graph $\g(k)=\g[W(k)]$.

\subsection{Time invariant dynamics}

In the special case of constant $W(k)= W$ for all $k$, consensus conditions are well-known and dual to the conditions of regularity in a stationary discrete-time Markov chain, discussed, e.g., in~\citep{ProTempo:2017-1}.
\begin{theorem}\label{thm.static-eq}
For $W(k)= W$, $k=0,1,\ldots$, where the constant matrix $W$ is row-stochastic, the following conditions are equivalent:
\begin{enumerate}
\item for any initial condition $x(0)$, the opinions $x_j(k)$ converge to some consensus value, that is
\be\label{eq.consen}
\quad\forall x(0)\,\,\exists c=c(x(0)):\quad x(k)\xrightarrow[k\to\infty]{}c\ones_n.
\ee
\item the matrix $W$ is SIA\footnote{SIA matrices are also called regular~\citep{Seneta} or fully regular~\citep{GantmacherVol2}, since they correspond to regular (ergodic) Markov chains~\cite{Seneta} that ``forget'' their history and converge to a unique stationary distribution. Matrix $W$ is regular if and only if some column of $W^k$ is strictly positive for some $k$~\citep{Seneta}.}, that is, there exists a nonnegative vector $\pi\in\r^n$ such that $\lim\limits_{k\to\infty}W^k=\ones_n\pi^{\top}$;
\item $\g[W]$ is a quasi-strongly connected graph whose (unique) source component  is aperiodic\footnote{We recall that a strong component of a graph is a source component if no arc enters it, a quasi-strongly connected graph has only one such component. A dual formulation in terms of Markov chains is: a chain is regular if and only if has only one essential (recurrent) class, which is aperiodic~\citep{GantmacherVol2,Seneta}.}.
\end{enumerate}
If these conditions hold, then $\pi$ is the Perron-Frobenius left eigenvector of $W$, such that $\pi^{\top}W=\pi^{\top}$ and $\sum_i\pi_i=1$. The consensus opinion of the group is $c=\pi^{\top}x(0)$.
\end{theorem}

{If the source component of the graph contains a set of nodes $I$, then the opinions of the agents from $I$ evolve independently of the remaining group:}
\[
w_{ij}=0\quad\forall i\in I,j\not\in I\Longrightarrow x_i(k+1)=\sum_{j\in I}w_{ij}x_j(k)\quad\forall i\in I.
\]
This explains the impossibility of consensus in presence of two such components. If a strong component $I$ is a source
and periodic, then the submatrix $(W_{ij})_{i,j\in I}$ has an eigenvalue $\la\in\mathbb{C}$ such that $\la\ne 1$ but $|\la|=1$, so that for almost all initial conditions the opinions periodically oscillate~\citep{ProTempo:2017-1}.

\subsection{Time-varying averaging: necessary conditions}
We now turn the attention to the more sophisticated case of non-stationary averaging procedure~\eqref{eq.degroot-classic}, with time-varying  $W(k)$. Notice first that consensus in the sense of definition~\eqref{eq.consen} can be established by ``degenerate'' procedures of iterative averaging. For instance, if $W(0)=\ones_n\pi^{\top}$ is a trivial rank-one stochastic matrix, then the opinion iteration terminates in one step: $x(1)=x(2)=\ldots=(\pi^{\top}x(0))\ones_n$. Similarly, if $W(k)W(k-1)\ldots W(0)=\ones_n\pi^{\top}$ is a rank-one matrix, the iterative averaging procedure terminates and establishes consensus in no more than $k$ steps. Notice that the dynamics of matrices $W(s)$, $s\ge k$ play no role: consensus remains invariant even if the agents do not communicate after the first $k$ steps. Although the problem of finite-time consensus is of self-standing interest~\citep{Hendrickx:2015}, the aforementioned situation is non-generic and it is usually ruled out  by a stronger requirement~\citep{Moro:05,ShiJohansson:13-1}
of establishing consensus for each starting time:
\be\label{eq.consen1}
\begin{aligned}
\forall k_0\ge 0,\,\forall x(k_0)\;\;\exists c=c(k_0,x(k_0)):\\
W(k)\ldots W(k_0) x(k_0)\xrightarrow[k\to\infty]{}c\ones_n.
\end{aligned}
\ee
As it has been already mentioned, consensus in~\eqref{eq.degroot-classic} is equivalent to \emph{strong ergodicity}~\citep{Seneta} of the backward matrix products $\mathcal W^{k_0,k}\dfb W(k)W(k-1)\ldots W(k_0)$, that is, the existence of a limit $\mathcal W^{k_0,\infty}=\lim_{k\to\infty}\mathcal W^{k_0,k}$, which is a one-rank stochastic matrix $\mathcal W^{k_0,\infty}=\ones_n\pi_{k_0}^{\top}$. To the best of the authors knowledge, no necessary and sufficient condition for consensus (or for strong ergodicity of backward products) has been obtained yet in the literature.
Many consensus criteria available in the literature are based on various \emph{ergodicity coefficients} (see e.g.~\citep{Dobrushin:56,Hajnal:58,Seneta,LEIZAROWITZ1992189,CaoMorse:08}) and so-called properties of infinite flow~\citep{TouriNedic:11,TOURI20121477,Bolouki:2016}.
In this review, we confine ourselves to a few criteria that seem to be most convenient as they admit simple graph-theoretic interpretations.

An important \emph{necessary} condition for consensus is the quasi-connectivity of the so-called \emph{persistent} graph~\citep{ShiJohansson:13-1,XiaShiCao2019}. In continuous time, this property is called integral or essential connectivity~\citep{MartinHendrickx:2016,MatvPro:2013}.
\begin{definition}
A \emph{persistent graph}\footnote{In many works on consensus~\cite{Blondel:05,Moro:05}, an additional assumption is stipulated that non-zero influence weights $w_{ij}$ are uniformly positive, see the condition~\eqref{eq.unipos}. Then $(j,i)\in\e_p$ if and only if the influence of $j$ on $i$ ``persists'' in the sense that arc $(j,i)$ appears in infinitely many graphs $\g[k]$.} of the matrix sequence $\{W(k)\}$ is the graph $\g_p=([1:n],\e_p)$, whose set of arcs is defined as follows
\be\label{eq.g-p}
\e_p=\left\{(j,i):\sum_{k=0}^{\infty}w_{ij}(k)=\infty\right\}.
\ee
\end{definition}

\begin{lemma}\cite[Proposition 3.1]{ShiJohansson:13-1}\label{lem.persist}
If consensus in the sense~\eqref{eq.consen1} is established, then the persistent graph $\g_p$ is quasi-strongly connected.
\end{lemma}

The necessary condition from Lemma~\ref{lem.persist} implies another necessary condition proposed by~\cite{TOURI20121477}
under the name of \emph{infinite flow property}: for every set of indices $S\subsetneq [1:n]$, $S\ne\emptyset$, the graph $\g_p$ contains an arc $(i,j)$ between $S$ and $S^c$, that is, either $i\in S,j\in S^c$ or $i\in S^c$ and $j\in S$. This is obvious, since either $S$ or $S^c$ should contain the spanning tree's root. At the same time,~\cite{TOURI20121477} establish a stronger necessary condition for consensus (ergodicity of the matrix products), which is called \emph{absolute flow property} and requires that the infinite flow property remains invariant under special ``rotational'' transformations of the matrices $W(k)$; an extension of this result obtained by~\citep{Bolouki:2016} states the necessity of a so-called \emph{infinite jet-flow property}. To verify these properties is, however, a self-standing non-trivial problem, which has been solved only in special cases~\citep{TOURI20121477,Bolouki:2016}.

In the case of static matrix $W(k)\equiv W$, $\forall k$, $\g_p=\g[W]$, the necessary condition from Lemma~\ref{lem.persist} becomes ``almost'' sufficient in view of Theorem~\ref{thm.static-eq} (modulo the aperiodicity assumption, which holds, e.g., if $w_{ii}>0\;\forall i$). For non-stationary matrices, even completeness of the graph $\g_p$ does not always imply consensus if some persistent interactions are much ``weaker'' than others and the divergence rates of the series in~\eqref{eq.g-p} are different~\citep{Moro:05}.

To guarantee consensus, the quasi-strong connectivity of $\g_p$ has to be supplemented by additional assumptions.
Three typical conditions of this type are ({\em i}) \emph{repeated} (uniform) connectivity,
({\em ii}) uniformly bounded ratios of influence weights on persistent arcs (``arc-balance''), and ({\em iii}) a uniform version of the cut-balance condition~\eqref{eq.cut-balance-static1}. We focus only on convergence to consensus and do not consider here additional properties of the algorithms such as, e.g., their convergence rates~\citep{Blondel:05,CaoMorse:08,OlshevskyTsitsi:11,XiaShiCao2019}.

\subsection{Sufficient conditions: repeated connectivity}

Conditions of the first type are known as the \emph{repeated} (periodic, uniform) quasi-strong connectivity. Typically, these conditions are formulated under additional assumption of uniform positivity of non-zero weights $w_{ij}(k)$. The following consensus criterion is known from~\citep[Theorem~1]{Blondel:05}, \citep[Theorem~2]{Moro:05} and \citep[Theorem 2.39]{RenBeardBook}.
\begin{lemma}\label{lem.repeated-qs}
Suppose that all non-zero entries of $W(k)$ are uniformly positive:
\be\label{eq.unipos}
w_{ij}(k)\in\{0\}\cup[\eta,1]\qquad\forall i,j\in[1:n]\,\forall k\ge 0,
\ee
furthermore, $w_{ii}(k)>0$ for any $i,k$. Assume also that a period $T>0$ exists such that the graphs
\[
\g_{k,T}=\g[W(k)+\ldots+W(k+T-1)]
\]
(arising as unions of $T$ consecutive graphs $\g[W(k)],\ldots,\g[W(k+T-1)]$) are quasi-strongly connected for all $k\ge 0$. Then consensus~\eqref{eq.consen1} is established.
\end{lemma}

The assumptions of Lemma~\ref{lem.repeated-qs} are known as the \emph{repeated quasi-strong connectivity} and, obviously,
imply the quasi-strong connectivity of the persistent graph $\mathcal\g_p$.

\subsection{Sufficient conditions: arc-balance and cut-balance}

Sufficient conditions of the second kind have been proposed in~\citep{ShiJohansson:13-1} under name of ``arc-balance''.
This condition requires that all persistent interactions occur simultaneously, moreover, the ratios of corresponding influence rates on \emph{every two persistent arcs} $(j,i),(m,l)\in\e_p$ is uniformly bounded
\be\label{eq.arc-balance}
C^{-1}w_{lm}(k)\le w_{ij}(k)\le Cw_{lm}(k)\;
\forall k\ge 0.
\ee
A ``non-instantaneous'' relaxation of this condition was introduced in~\citep{XiaShiCao2019}: for each pair of persistent arcs $(j,i),(m,l)\in\e_p$ and time instant $k_0\ge 0$ one has
\be\label{eq.arc-balance1}
C^{-1}\sum_{k=k_0}^{k_0+L}w_{lm}(k)\le\sum_{k=k_0}^{k_0+L}w_{ij}(k)\le C\sum_{k=k_0}^{k_0+L}w_{lm}(k).
\ee
Here $C\ge 1, L\ge 0$ are, respectively, a real and an integer numbers that are independent of $(j,i)$, $(m,l)$ and $k_0$.

\begin{lemma}\label{lem.arc-balance}\citep{XiaShiCao2019}
Let the diagonal entries of $W(k)$ be uniformly positive $w_{ii}(k)\ge\eta>0\,\forall i,k$ and~\eqref{eq.arc-balance1} hold for some $C\ge 1,L\ge 0$. Then consensus~\eqref{eq.consen1} is established if and only if $\g_p$ is quasi-strongly connected.
\end{lemma}

An alternative consensus condition, introduced in~\citep{TsiTsi:13}, and, later in~\citep{Bolouki2015,MartinHendrickx:2016} is the \emph{uniform cut-balance}: for any cut $(I,J)$ and any $k$,
\be\label{eq.cut-balance-dynam}
\sum_{i\in I,j\in J}w_{ij}(k)\le C\sum_{i\in I,j\in J}w_{ji}(k),
\ee
where constant $C$ is independent of $k$ and of the cut (for static graphs, this condition coincides with~\eqref{eq.cut-balance-static}). A more general condition introduced in~\citep{XiaShiCao2019} replaces~\eqref{eq.cut-balance-dynam} by its ``non-instantaneous'' version:
for any cut $(I,J)$ and any $k_0\geq 0$ one has
\be\label{eq.cut-balance-dynam1}
\sum_{k=k_0}^{k_0+L}\sum_{i\in I,j\in J}w_{ij}(k)\le C\sum_{k=k_0}^{k_0+L}\sum_{i\in I,j\in J}w_{ji}(k),
\ee
where a real number $C\ge 1$ and an integer $L\ge 0$ are independent of the choice of $(I,J)$ and $k_0$.

\begin{lemma}\label{lem.cut-balance}\citep{XiaShiCao2019}
Let the diagonal entries of $W(k)$ be uniformly positive $w_{ii}(k)\ge\eta>0\,\forall i,k$ and~\eqref{eq.cut-balance-dynam1} hold for some $C\ge 1,L\ge 0$. Then consensus~\eqref{eq.consen1} is established if and only if graph $\g_p$ is quasi-strongly connected\footnote{Under the assumption~\eqref{eq.cut-balance-dynam1}, the graph $\g_p$ has isolated strong components, so quasi-strong and strong connectivity are equivalent.}.
\end{lemma}

\section{New results: convergence of RAIs}\label{sec.main}
In this section, we present novel results concerned with the behavior of any feasible solution to the RAI~\eqref{eq.ineq}. The proofs of these results will be given in Section~\ref{sec.proofs}.
The RAI constitutes a relaxation of the averaging equation \eqref{eq.degroot-classic}, and clearly any trajectory of \eqref{eq.degroot-classic} is also
a feasible trajectory for \eqref{eq.ineq}. Hence, any result that holds for all feasible  trajectories of a RAI also holds for the trajectories of \eqref{eq.degroot-classic}.
 We next develop a theory for the RAI model, and then
discuss  in Section~\ref{sec.appl} several relevant applications.
We are primarily interested in the properties of convergence and consensus, as formalized in the following
\begin{definition}\
\begin{enumerate}
\item The RAI~\eqref{eq.ineq} is \emph{convergent} if
all of its feasible solutions converge, that is, for any feasible $\{x(k)\}$
there exist  $\bar x$ such that
\be\label{eq.limit}
\lim_{k\to\infty} x(k) = \bar x.
\ee
\item
The RAI~\eqref{eq.ineq} establishes \emph{consensus} if it is convergent and
the terminal opinions are coincident $\bar x_1=\ldots=\bar x_n$ (equivalently, $\bar x=c\mathbbm{1}_n$, where $c$ is a scalar).
\end{enumerate}
\end{definition}
Obviously, $x(k)$ is a feasible solution to~\eqref{eq.ineq} if and only if
\be\label{eq.delta}
\Delta(k)=W(k)x(k)-x(k+1)\ge 0.
\ee
A feasible solution to the RAI thus can be considered as a solution for the ``forced'' recursion
\[
x(k+1)=W(k)x(k)-\Delta(k),
\]
where the forcing term  $\Delta(k)\ge 0$ may depend on the trajectory, be unknown and unbounded.
In spite of the input-to-state stability of consensus algorithms~\citep{ShiJohansson:13}, the exact consensus~\eqref{eq.consen1} is destroyed by an arbitrarily small bounded disturbance.
The capability of RAI~\eqref{eq.ineq} to establish consensus can be thus considered as a counterintuitive
robustness property of the associated iterative averaging procedure~\eqref{eq.degroot-classic} against
\emph{unbounded yet sign-preserving} disturbances.

Some components $\bar x_i$ of the limit vector in~\eqref{eq.limit} may be equal to $-\infty$, since~\eqref{eq.ineq}
does not guarantee the existence of a finite lower bound for the solution;
however, the solution always has a finite \emph{upper} bound as implied by the following straightforward proposition.
\begin{proposition}\label{prop.bound}
If $x(k)$ is a feasible solution for the RAI~\eqref{eq.ineq}, then the sequence $M(k)=\max_ix_i(k)$ is non-increasing
$M(k+1)\leq M(k)\leq\ldots\leq M(0)$.
\end{proposition}

Unlike the monotone sequence $M(k)$, the behavior of the minimum opinion $m(k)=\min_i x_i(k)$ and the opinions'
``diameter'' $d(k)=M(k)-m(k)$ can be very different. Even for bounded solutions, the diameter's non-increasing property
fails, and the diameter cannot serve as a natural Lyapunov function. This first principal difference
between the RAI~\eqref{eq.ineq} and the classical iterative averaging procedure~\eqref{eq.degroot-classic} makes
inapplicable the plethora of methods based, explicitly or implicitly, on the ``pseudo-contracting'' properties of
iterative averaging models~\citep{Fang:08}, and estimates of the diameter $d(k)$~\citep{Blondel:05,Moro:05,ShiJohansson:13-1,XiaShiCao2019}
cannot be used to prove convergence to consensus. Another principal difference between RAI~\eqref{eq.ineq}
and the averaging procedure~\eqref{eq.degroot-classic} is the absence of duality between consensus and ergodicity of
backward matrix products. Iterating~\eqref{eq.ineq}, one easily derives the one-sided inequality
$
x(k+1)\leq W(k)\ldots W(0)x(0),
$
however, the convergence of the right-hand side says nothing about the behavior of the solution.
This is not surprising since, as it will be shown, the strong ergodicity
of backward products $W(k)\ldots W(0)$ (equivalently, consensus in~\eqref{eq.degroot-classic})
is \emph{not} sufficient for the convergence of solutions in~\eqref{eq.ineq}. For this reason, methods
based on matrix analysis and Markov chain theory~\citep{Seneta,RenBeardBook,CaoMorse:08,TOURI20121477,Bolouki:2016} are also inapplicable to analysis of the RAI.


Notice that all results presented in this section are applicable (with straightforward modifications) to the RAI
\[
y(k+1)\ge W(k)y(k)
\]
that reduce to~\eqref{eq.ineq} by the transform $y(k)\mapsto x(k)=-y(k)$.


\subsection{Time-invariant RAI: convergence and consensus}

We start with our first result, which is a counterpart of Theorem~\ref{thm.static-eq}.

\begin{theorem}\label{thm.static-ineq}
The RAI~\eqref{eq.ineq} with $W(k)\equiv W$, $\forall\, k$, where $W$ is a row-stochastic matrix, is convergent
if and only if all strong components of $\g[W]$ are isolated and aperiodic. If this condition holds, then
\begin{enumerate}
\item $\bar x_i=\bar x_j$ whenever $i$ and $j$ belong to the same strongly connected component of the graph $\g[W]$ (``partial consensus'' is established);
\item if the opinion $x_i(k)$ converges to a finite limit, the corresponding residual vanishes, i.e.,
\be\label{eq.resid0}
\Delta_i(k)\xrightarrow[k\to\infty]{}0.
\ee
\end{enumerate}
The RAI establishes consensus if and only if $\g[W]$ is a strongly connected aperiodic graph, that is, $W$ is a \emph{primitive} (irreducible aperiodic) matrix.
\end{theorem}

Comparing Theorems~\ref{thm.static-eq} and~\ref{thm.static-ineq}, one notes that consensus in the RAI requires the \emph{strong} connectivity, whereas inequalities over quasi-strongly connected graphs cannot ensure convergence of each solution. For instance, the RAI
\[
x_1(k+1)\le x_1(k),\quad x_2(k+1)\le\frac{x_1(k)+x_2(k)}{2},
\]
has a non-converging solution $x_1(k)\equiv 1$, $x_2(k)=(-1)^k$.

\subsection{Time-varying case: convergence under reciprocal interactions}

Convergence of the solutions in the time-varying case will be established under several assumptions.
Our first assumption, which is typically adopted in the works on consensus~\citep{Blondel:05,RenBeardBook,ShiJohansson:13-1,XiaShiCao2019}, can be considered as a counterpart of the aperiodicity assumption. Theorem~\ref{thm.static-ineq} shows that, even for the static graph case, this assumption can be relaxed yet not discarded completely.
\begin{assum}(\textbf{Self-influence})\label{ass.aperiod}
The diagonal entries of $W(k)$ are uniformly positive: a constant $\eta>0$ exists such that $w_{ii}(k)\ge\eta>0\,\forall i,k$.
\end{assum}
The second assumption requires uniform positivity of non-zero weights $w_{ij}$. This assumptions holds in many interesting applications (see Section~\ref{sec.appl}) and allows to simplify the proofs. At the same time, this assumption is in fact not necessary for convergence and consensus and can be dropped in some situations (see Theorem~\ref{thm.new} below).

\begin{assum}\label{ass.unipos}
The entries of $W(k)$ satisfy the condition~\eqref{eq.unipos}, i.e., all non-zero entries are uniformly positive.
\end{assum}

\begin{remark}\label{rem.unipos}
Under Assumption~\ref{ass.unipos}, persistent arc $(j,i)\in\e_p$ stands for a pair of agents that interacts infinitely often, i.e. $w_{ij}(k_s)>0$ for an infinite sequence $k_s\to\infty$.
\end{remark}

We start with a counterpart of Lemma~\ref{lem.persist}, which gives a \emph{necessary} condition for consensus and shows that the results of Lemmas~\ref{lem.repeated-qs}
and~\ref{lem.arc-balance} do not retain their validity for the inequalities. Even for repeated quasi-strong connectivity, some solutions to RAI~\eqref{eq.ineq} may fail to converge.
\begin{lemma}\label{lem.persist1}
Under Assumption~\ref{ass.unipos}, RAI~\eqref{eq.ineq} can be convergent (respectively, establishes consensus) only if
all strong components of the graph $\g_p$ are isolated (respectively, $\g_p$ is a \emph{strongly connected} graph).
\end{lemma}
As it has been discussed already, this necessary condition is not sufficient for consensus, even for the classical iterative averaging procedure~\eqref{eq.degroot-classic}, see, e.g., the counterexample in~\cite{Moro:05}.
To ensure consensus, some additional assumption of \emph{reciprocity} of the interactions is needed. We now introduce such an assumption, whose detailed discussion will be given in Section~\ref{subsec.recipro}.

\begin{definition}
For two non-empty subsets $I,J\subseteq[1:n]$, let $a_{I,J}(k_0:k_1)$ denote \emph{the number of arcs} connecting $J$ to $I$ over the time window $[k_0:k_1]$, that is,
\[
a_{I,J}(k_0:k_1)\dfb\left|\{(i,j):w_{ij}(k)>0\;\;\text{for some $k\in[k_0:k_1]$}\}\right|.
\]
\end{definition}


\begin{assum}(\textbf{Reciprocity})\label{ass.recipro}
There exist integer numbers $\M\ge 1,T\ge 0$ such that for any cut $(I,J)$ the following implication holds:
\be\label{eq.gen-balance}
a_{I,J}(k_0:k_1)\ge \M\Longrightarrow a_{J,I}(k_0:(k_1+T))\ge 1.
\ee
Thus a sufficiently large cumulative influence of group $J$ onto group $I$ during some time interval $[k_0:k_1]$
triggers the response from $I$ to $J$ (possibly, retarded by $T$ steps).
\end{assum}

\noindent
We are now in position to formulate our second result.
\begin{theorem}\label{thm.dynamic-ineq}
Let Assumptions~\ref{ass.aperiod}~,\ref{ass.unipos} and~\ref{ass.recipro} hold. Then
\begin{enumerate}
\item The RAI~\eqref{eq.ineq} is convergent;
\item the terminal opinions of the agents $i,j\in[1:n]$ from the same strong component of $\g_p$ coincide $\bar x_i=\bar x_j$ (i.e., partial consensus is established);
\item if the opinion $x_i(k)$ converges to a finite limit, then the corresponding residual vanishes $\Delta_i(k)\xrightarrow[k\to\infty]{}0$.
\end{enumerate}
The RAI~\eqref{eq.ineq} establishes consensus if and only if $\g_p$ is strong.
\end{theorem}

Theorem~\ref{thm.dynamic-ineq} can be further extended: both convergence and consensus appear to be robust against bounded \emph{communication delays}. For a classical consensus algorithm, such a robustness is known under the assumption of repeated quasi-strong connectivity~\citep{Bliman:06}.
\begin{theorem}\label{thm.robust}
Consider a sequence of matrices $(d_{ij}(k))_{i,j=1}^n$, whose diagonal entries are zero\footnote{In other words, an agent has access to its own \emph{undelayed} state.} $d_{ii}(k)=0$, the other entries being uniformly bounded $0\le d_{ij}(k)\leq d_*<\infty$. 
Then the criteria from Theorem~\ref{thm.dynamic-ineq} retain their validity for the inequalities
\be\label{eq.ineq-d}
x_i(k+1)\le \sum_{j=1}^nw_{ij}(k)x_j(k-d_{ij}(k)),\quad i\in[1:n].
\ee
Namely, the delayed RAI~\eqref{eq.ineq-d} is convergent if Assumptions~\ref{ass.aperiod},\ref{ass.unipos} and~\ref{ass.recipro} hold. Also, if the component $x_i(k)$ is bounded, then the corresponding ``residual'' vanishes
\[
\Delta_i(k)=\sum_{j=1}^nw_{ij}(k)x_j(k-d_{ij}(k))-x_i(k+1)\xrightarrow[k\to\infty]{}0\;\forall i.
\]
If, additionally, the graph $\g_p$ is strongly connected, then the RAI establishes consensus.

If both weights $w_{ij}$ and delays $d_{ij}$ are \emph{constant}, Assumption~\ref{ass.aperiod} can be relaxed:
it suffices that each strongly connected component of $\g[W]$ contains a self-arc (standing for the diagonal entry $w_{ii}>0$).
\end{theorem}

Notice that the result of Theorem~\ref{thm.static-ineq}, replacing Assumption~\ref{ass.aperiod} by \emph{aperiodicity}
of each strong component of $\g[W]$, does not retain its validity even if the delays are constant.
Such a robustness cannot be proved even for the algorithm~\eqref{eq.degroot-classic}.
A trivial counterexample is
\be\label{eq.counter-ex1}
\begin{gathered}
x_1(k+1)=\frac{x_2(k-1)+x_3(k)}{2},\\
x_2(k+1)=x_1(k-1),\quad x_3(k+1)=x_2(k).
\end{gathered}
\ee
The corresponding graph $\g[W]$ is strongly connected and aperiodic since it
contains two cycles $1\xrightarrow[]{} 2\xrightarrow[]{}1$ (length 2) and $1\xrightarrow[]{} 2\xrightarrow[]{}3\xrightarrow[]{}1$ (length 3). At the same time, the system obviously has infinitely many
periodic solutions: choosing $x_1(k)$ to be a sequence of period $4$, the vector $(x_1(k),x_1(k-2),x_1(k-3))$ is a solution to~\eqref{eq.counter-ex1}.

In presence of time-varying delays, the strong connectivity and positivity of a single diagonal element $w_{ii}$ can be insufficient, as shown by the following example. Let $n=2$ and $w_{11}=0,w_{12}=1,w_{21}=w_{22}=1/2$. The graph $\g[W]$ is then strongly connected and aperiodic since node 2 has a self-loop. Choosing the delays $d_{11}=d_{22}=d_{12}=0$ and $d_{21}(k)=k\mod 2\in\{0,1\}$, the sequence
\[
x_1(k)
=\begin{cases}
1, & k\;\;\text{odd}\\
0, & k\;\;\text{even}
\end{cases},\quad x_2(k)\equiv 1
\]
satisfies the RAI~\eqref{eq.ineq-d}. Indeed, $x_1(k)\le x_2(k)=w_{11}x_1(k)+w_{12}x_2(k)$. Also, $x_1(k-d_{21}(k))=1$ for any $k$ and hence
$x_2(k)=w_{21}x_1(k)+w_{22}x_2(k)$. Therefore, the delayed RAI~\eqref{eq.ineq-d} fails to be convergent.

\subsection{Reciprocity condition: discussion}\label{subsec.recipro}

In this subsection, we discuss the relation between our results and previously known consensus criteria.

\subsubsection{Consensus under repeated strong connectivity}

As it has been already discussed, Lemma~\ref{lem.repeated-qs} does not retain its validity for the RAI,
even for constant $W(k)\equiv W$. Notice, however, that if all graphs $G[W(k)+\ldots+W(k+T-1)]$, $k\ge 0$ (equivalently,
unions of $T$ consecutive graphs) are \emph{strongly} connected, then
the implication~\eqref{eq.gen-balance} holds with an arbitrary $\mathfrak{M}\geq 0$,
since the statement on its right-hand side is always true
(during $T$ consecutive steps, some agent from $I$ has to communicate to some agent from $I$).
Theorem~\ref{thm.dynamic-ineq} implies the following counterpart of Lemma~\ref{lem.repeated-qs}, which has been reported in our previous work~\citep{ProCao2017-3}.

\begin{corollary}\label{cor.repeated}
Assume that Assumptions~\ref{ass.aperiod} and~\ref{ass.unipos} hold and a period $T>0$ exists such that all graphs $G[W(k)+\ldots+W(k+T-1)]$, $k\ge 0$ (equivalently, unions of $T$ consecutive graphs) are \emph{strongly} connected. Then the RAI~\eqref{eq.ineq} establishes consensus and~\eqref{eq.resid0} holds for any bounded solution.
\end{corollary}


\subsubsection{Non-instantaneous type-symmetry}

Another example where the assumptions of Theorem~\ref{thm.dynamic-ineq} hold is \emph{type-symmetry}~\citep{Lorenz:2005,TsiTsi:13}
of the interaction weights
\be\label{eq.type-symm}
w_{ij}(k)\le Cw_{ji}(k)\quad\forall i,j\in [1:n]\,\forall k\ge 0,
\ee
where $C\ge 1$ is a constant. Under Assumption~\ref{ass.unipos}, the type-symmetry can be reformulated as a condition of \emph{bidirectional} communication: if $j$ communicates to $i$ at time $k$ (that is, $w_{ij}(k)>0$), then $i$ communicates to $j$ (i.e. $w_{ji}(k)>0$). A natural extension~\citep{Blondel:05} of the latter condition allows a \emph{delayed} response:
\be\label{eq.type-symm-del}
\forall k_0\ge 0\quad w_{ij}(k_0)>0\Longrightarrow \sum_{k=k_0}^{k_0+T}w_{ji}(k)>0.
\ee
Condition~\eqref{eq.type-symm}, as well as~\eqref{eq.type-symm-del}, obviously implies
Assumption~\ref{ass.recipro} with $\mathfrak{M}=1$. Theorem~\ref{thm.dynamic-ineq} thus extends the result on convergence of type-symmetric consensus algorithms~\eqref{eq.degroot-classic}~\cite[Theorem~2]{Lorenz:2005},~\cite[Theorem~5]{Blondel:05} to the systems of RAI~\eqref{eq.ineq}.

\subsubsection{Uniform cut-balance and arc-balance}

Under Assumption~\ref{ass.unipos}, the condition of uniform cut-balance~\eqref{eq.cut-balance-dynam} implies
the validity of Assumption~\ref{ass.recipro} (with $\mathfrak{M}=1$, $T=0$). The more general condition~\eqref{eq.cut-balance-dynam1}
also implies the validity of Assumption~\ref{ass.recipro} with $\mathfrak{M}=1$, $T=L$.
Indeed, suppose that $a_{I,J}(k_0:k_1)\ge 1$ and let $k\in [k_0:k_1]$ be the first instant when $a_{I,J}(k_1')>0$.
Then~\eqref{eq.cut-balance-dynam1} implies that $a_{J,I}(k:(k+L))>0$, which proves~\eqref{eq.gen-balance}
with $T=L$.

In view of Remark~\ref{rem.unipos}, under Assumption~\ref{ass.unipos} the condition of ``arc-balance''~\eqref{eq.arc-balance}
implies that, starting from some instant $k_*$ one has either $\e[W(k)]=\emptyset$ or $\e[W(k)]=\e_p$. This obviously
implies $(1,0)$-reciprocity of the sequence $\{W(k)\}_{k\ge k_*}$. Similarly, the non-instantaneous
arc-balance~\eqref{eq.arc-balance1} implies that for $k\ge k_*$ either $\e[W(k)+\ldots+W(k+L)]=\emptyset$ or
$\e[W(k)+\ldots+W(k+L)]=\e_p$. Similar to the uniform cut-balance case, one
shows that Assumption~\ref{ass.recipro} holds with $\mathfrak{M}=1$, $T=L$.

Theorem~\ref{thm.dynamic-ineq} yields in the following counterpart of Lemmas~\ref{lem.arc-balance} and~\ref{lem.cut-balance}.

\begin{corollary}
Let Assumptions~\ref{ass.aperiod},~\ref{ass.unipos} be valid and one of the conditions~\eqref{eq.arc-balance1} or~\eqref{eq.cut-balance-dynam1} hold. Then the RAI~\eqref{eq.ineq}
is convergent and consensus in each strong component of $\g_p$ is established and~\eqref{eq.resid0} holds for any bounded solution. The RAI establishes consensus if and only if $\g_p$ is strong.
\end{corollary}


\subsubsection{Periodic gossiping with intermittent communication}

We now propose a simple example where the implication~\eqref{eq.gen-balance}
does not reduce to any of the aforementioned type-symmetry or balance conditions. A special class of iterative averaging policies~\eqref{eq.degroot-classic} is constituted by so-called \emph{gossiping} algorithms, where at each stage of the iteration at least one pair of agents communicates. In the case of deterministic  unidirectional \emph{periodic} gossip~\citep{Anderson2010} the sequence of graphs $\g[W(k)]$ is periodic and each graph $\g[W(k)]$ contains a single arc $\e[W(k)]=\{(j_k,i_k)\}$. Suppose now that these gossiping interactions are separated by arbitrarily long periods of ``silence'' where the agents do not interact. Formally, suppose that there exists a sequence $k_1<k_2<\ldots<k_s<\ldots$ and a \emph{periodic} sequence of arcs $(j_s,i_s)$ such that
\be\label{eq.gossip}
W(k)=
\begin{cases}
(1-\alpha_s)\mathbf{e}_{i_s}\mathbf{e}_{j_s}^{\top}+\alpha_sI_n,& k=k_s\\
I_n,& k\ne k_s\,\forall s.
\end{cases}
\ee
For the iterative averaging algorithm~\eqref{eq.degroot-classic}, the periods of silence do not change the asymptotic
behavior of the solution (except for its convergence rate) since the vector of opinions $x(k)$, obviously,
remains unchanged for $k=k_s+1,\ldots, k_{s+1}-1$. The RAI allows the opinion vector to evolve during the
silence periods (the the only restriction is the inequality $x(k+1)\le x(k)$), so the convergence is not straightforward.

All the aforementioned reciprocity conditions (type-symmetry, arc-balance and cut-balance) fail
to hold if $\sup_s(k_{s+1}-k_s)=\infty$. At the same time, the sequence of matrices~\eqref{eq.gossip}
satisfies the condition~\eqref{eq.gen-balance}, where $T=0$ and
$\M$ is the period of the sequence $(j_s,i_s)$. If $a_{I,J}(t_0:t_1)\ge \M$ for some time window $[t_0:t_1]$,
then during this time window each arc of the strongly connected graph $\g_p$  has appeared at least once,
and hence $a_{J,I}(t_0:t_1)>0$. Theorem~\ref{thm.dynamic-ineq} now implies the following.
\begin{corollary}\label{cor.gossip}
Let $W(k)$ have the representation~\eqref{eq.gossip}, where $(j_s,i_s)$ is a periodic sequence of arcs constituting
a strongly connected graph $\g_p$ and constants $\alpha_s$ obey the inequality $\eta\le\alpha_s\le 1-\eta$ for some $\eta>0$.
Then the RAI~\eqref{eq.ineq} establishes consensus and the residual vanishes~\eqref{eq.resid0} for any bounded component.
\end{corollary}

\subsection{Releasing Assumption~\ref{ass.unipos}}

As we have mentioned, Assumption~\ref{ass.unipos} is, in fact, \emph{not} necessary for convergence and consensus and can be discarded in some situations. In particular, this assumption can be dropped in the case of uniform cut-balance~\eqref{eq.cut-balance-dynam} and absence of delays.

\begin{theorem}\label{thm.new}
Assume that the graph $\g[W(k)]$ is uniformly cut-balanced~\eqref{eq.cut-balance-dynam} and Assumption~\ref{ass.aperiod} holds. Then all statements of Theorem~\ref{thm.dynamic-ineq} retain their validity, furthermore, if $x_i(k)$ is a bounded sequence, then $\sum_{k=0}^{\infty}\Delta_i(k)<\infty$.
\end{theorem}

Notice that Theorem~\ref{thm.new} for the case of iterative averaging procedure~\eqref{eq.degroot-classic} was first proved (in a more general stochastic formulation) in~\citep{TouriNedic:11,Touri:14}.

\section{Applications}\label{sec.appl}
In this section, we apply the main results on RAI to the analysis of several agent-based models of opinion formation and distributed algorithms.

\subsection{Stability of some positive delay systems}

A non-negative $n\times n$ matrix $A=(a_{ij})$ is \emph{substochastic} if $\sum_{j=1}^na_{ij}\leq 1\,\forall i$.
Unlike a stochastic matrix, always having eigenvalue at $1$, a substochastic matrix may be Schur stable $\rho(A)<1$. Theorem~\ref{thm.static-ineq} leads to an elegant stability
criterion~\citep{FrascaTempo:2013,Parsegov2015CDC}.
\begin{lemma}\label{lem.substoch}
For a substochastic matrix $A$, consider the set of ``deficiency'' indices $I_d=\{i:\sum_ja_{ij}<1\}$.
If every node $i$ in the graph $\g[A]$ is reachable from the set $I_d$ by a walk (formally, $i$ is reachable from at least one node $j\in I_d$) then $\rho(A)<1$. In particular, if $A$ is irreducible ($\g[A]$ is strongly connected) and $I_d\ne\emptyset$, then $\rho(A)<1$.

More generally, if a constant matrix $(d_{ij})_{i,j=1}^n$ has zero diagonal entries $d_{ii}=0$, then the following linear delay system is globally asymptotically stable 
\be\label{eq.delay-syst}
x_i(k+1)=\sum_{j=1}^na_{ij}x_j(k-d_{ij}),\quad i\in[1:n].
\ee
\end{lemma}
\begin{proof}

We are going to show that every solution $x(k)$, $k=0,1,\ldots$, of~\eqref{eq.delay-syst} obeys the RAI~\eqref{eq.ineq-d}, with a special constant row-stochastic matrix $W$ ``dominating'' the matrix $A$ in the sense that $w_{ij}\ge a_{ij}$ for all $i,j$.

Consider first the undelayed case, where $x(k)=A^kx(0)$ for every $k$. Let $V=\diag(A\ones_n)$ be the diagonal matrix, whose entries $v_{ii}=\sum_{l=1}^na_{il}\leq 1$ stand for the sums of matrix $A$'s rows.
Obviously, the matrix
\be\label{eq.stoch-aux}
W=A+\frac{1}{n}(I-V)\ones_n\ones_n^{\top}
\ee
is stochastic since its entries are non-negative and $W\ones_n=A\ones_n+(I-V)\ones_n=\ones_n$. Notice also that
$w_{ij}>a_{ij}\ge 0\,\forall j$ for every $i\in I_d$. Hence in the graph $\g[W]$ each node $j$ is connected to every node from $I_d$;
in particular, all nodes from $I_d$ have self-loops. By assumption, from $I_d$ every node is reachable by a walk, therefore $\g[W]$ is strongly connected. Choosing an arbitrary non-negative vector $x_0\ge 0$, the vectors $x(k)=A^kx_0$ are non-negative for any $k\ge 0$ and satisfy the inequality~\eqref{eq.ineq} with
$W(k)\equiv W$. Thanks to Theorem~\ref{thm.static-ineq}, $x(k)\xrightarrow[]{} c\ones$, where $c\ge 0$ and
\be\label{eq.aux2+}
\Delta(k)=Wx(k)-x(k+1)=n^{-1}(I-V)\ones_n\ones_n^{\top}x(k)\xrightarrow[k\to\infty]{}0.
\ee
The latter condition implies that $c(I-V)\ones_n=0$, which is only possible for $c=0$ (by assumption, $I_d(A)\ne\emptyset$).
Hence $A^kx_0\xrightarrow[k\to \infty]{}0$ for any vector $x_0\ge 0$. Since every vector is a difference of two non-negative vectors, $A$ is Schur stable.

The second statement is proved similarly. Let $W$ be as in~\eqref{eq.stoch-aux}, and
consider a solution to~\eqref{eq.delay-syst} with non-negative initial condition $x_i(\tau)\ge 0$ for $-d_*\le \tau\le 0$. This solution is non-negative and obeys the inequalities
\[
0\le x_i(k+1)\leq\sum_{j=1}^nw_{ij}x_j(k-d_{ij})\quad\forall i\in [1:n].
\]
Theorem~\ref{thm.robust} ensures consensus $x(k)\xrightarrow[k\to\infty]{} c\ones$ and
\be\label{eq.aux2}
\begin{aligned}
\Delta_i(k)=\sum_{j=1}^nw_{ij}x_j(k-d_{ij})-x_i(k+1)=\\=
n^{-1}(1-v_{ii})\sum_{i=1}^nx(k)\xrightarrow[k\to\infty]{}0,
\end{aligned}
\ee
i.e., $c(1-v_{ii})=0\,\forall i$ and thus $c=0$. Hence, solutions with non-negative initial conditions vanish as $k\to\infty$. Due to the linearity of~\eqref{eq.delay-syst}, the same holds for an arbitrary solution, that is, the system is asymptotically stable.\epf
\end{proof}

The Schur stability criterion from Lemma~\ref{lem.substoch} is not only sufficient but also necessary~\citep{Parsegov2017TAC}. Lemma~\ref{lem.substoch} implies, in particular, the condition of opinion convergence in the \emph{Friedkin-Johnsen} model of opinion formation~\citep{FrascaTempo:2013}. To obtain an explicit estimate of the $\rho(A)$ (that is, the solution's convergence rate) in terms of the weighted graph $\g[A]$ is a non-trivial problem; some results are available in~\cite{ProTempoCao16-2}.

\subsection{The Hegselmann-Krause model with ``truth seekers''}

One of the seminal models describing opinion formation in social networks is known as the Hegselmann-Krause \emph{bounded confidence} model. Its simplest version~\citep{Krause,Krause:2002} arises as a modification of the DeGroot algorithm~\eqref{eq.degroot-classic}, where the matrix $W(k)$ co-evolves with the opinion vector and is defined as follows
\be\label{eq.hk-1}
\begin{gathered}
W(k)=\bar W(x(k)),\;\bar w_{ij}(x)\dfb
\begin{cases}
\frac{1}{|N_i(x)|},&j\in N_i(x),\\
0,&j\not\in N_i(x),
\end{cases}\\
N_i(x)\dfb\{j:|x_j-x_i|<\ve\}.
\end{gathered}
\ee
In other words, at each stage of the opinion iteration, an individual replaces his/her 
opinion by the average of its own ($N_i(x)\ni i$) opinion and the opinions of \emph{like-minded} 
individuals that belong to the \emph{confidence set}, that is, the ball $B_{\ve}(x_i)=\{x: |x-x_i|<\ve\}$. 
Agents ignore dissimilar opinions, i.e., those lying outside their confidence sets.

Notice that the matrix $\bar W(x)$ is stochastic and satisfies the condition~\eqref{eq.unipos} with $\eta=n^{-1}$ since $1\le |N_i(x)|\le n\,\forall i\in[1:n]$. Also, $\bar w_{ii}(x)\ge n^{-1}\,\forall i$. In particular, the sequence of matrices $W(k)=\bar W(x(k))$ satisfies the \emph{type-symmetry} condition~\eqref{eq.type-symm} with $C=n$. The corresponding graph $\g_p$ is thus undirected and $(j,i)\in\e_p$ if and only if $w_{ij}(k)>0$ (equivalently, $|x_i(k)-x_j(k)|<\ve$) for infinitely many $k\ge 0$. One thus arrives at a simple proposition.
\begin{proposition}\label{prop.aux}
For any sequence of vectors $x(k)$, the sequence $W(k)=\bar W(x(k))$ satisfies Assumptions~\ref{ass.aperiod},\ref{ass.unipos} and~\ref{ass.recipro}.
\end{proposition}

In view of Proposition~\eqref{prop.aux} and Theorem~\ref{thm.dynamic-ineq} every solution to the Hegselmann-Krause model~\eqref{eq.degroot-classic},\eqref{eq.hk-1} converges.\footnote{For a group of $n$ agents, the opinion evolution actually terminates in $O(n^3)$ steps for the case of scalar opinions and $O(n^4)$ in the case of multidimensional opinions, see~\citep{ProTempo:2018} for a historical survey of the relevant results.} Since $(j,i)\not\in\e_p$ if and only if
$|x_i(k)-x_j(k)|\ge\ve$ for large $k$, the steady opinions $\bar x_i=\lim x_i(k)$ and $\bar x_j=\lim x_j(k)$ are either coincident or sufficiently distant $|\bar x_i-\bar x_j|\ge\ve$.

In this subsection, we consider a more general model with \emph{``truth seekers''}~\citep{HegselmannKrause:2006}
\be\label{eq.hk-2}
x(k+1)=(I-A) \bar W(x(k))x(k)+tA\ones_n.
\ee
Here $A=\diag(a_{11},\ldots,a_{nn})$ is a diagonal matrix, $0\le a_{ii}\le 1$, $\bar W(x)$ is defined in~\eqref{eq.hk-1} and
$t\in\r$ is a constant variable, referred to as \emph{truth} value. The presence of an additional term in the right-hand side of~\eqref{eq.hk-2} is explained by availability of some information to the agents, making their opinions closer to the truth value. The coefficient $a_{ii}$ measure the level of agent $i$'s ``awareness'' of the truth value~\citep{HegselmannKrause:2006}.
If $a_{ii}=1$, the agent is able to find the truth at a single step without communicating to the other agents. If $a_{ii}=0$, agent $i$ updates his/her opinion in accordance with the usual Hegselmann-Krause model, and the dynamics of his/her opinions are not (directly) influenced by the truth.
If $0<a_{ii}<1$, the opinion of individual $i$ is driven by both the truth and the opinions of the other individuals.

The presence of the static term in the right-hand side of~\eqref{eq.hk-2} visibly changes the dynamics and makes analysis of the model sophisticated. In the original work~\citep{HegselmannKrause:2006} it was shown that if $a_{ii}>0\,\forall i$, then all opinions converge to the truth $x(k)\xrightarrow[k\to\infty]{}t\ones_n$. Later~\cite{KurzRambau:2011} showed that
the opinions of the \emph{truth-seekers} ($a_{ii}>0$) always converge to $t$ even if the group has \emph{ignorant} agents (with $a_{ii}=0$). 
At the same time, the convergence of the ignorant agents' opinions remained an open problem, which has been solved in~\cite{Chazelle:11} 
by using a method of power series (``s-energy''). The aforementioned results employ diverse and highly non-trivial mathematical techniques. Theorem~\ref{thm.dynamic-ineq} allows to examine convergence properties of~\eqref{eq.hk-2} in a simpler way, as formalized 
in the following theorem.

\begin{theorem}\label{thm.hk}
If $a_{ii}>0$ for some $i$, then the following statements hold:
\begin{enumerate}
\item the limit $\bar x_i=\lim_{k\to\infty}x_i(k)$ exists for any $i$;
\item $\bar x_i=t$ if and only if agent $i$ persistently interacts with some truth-seeker: there exists $j$ such that $a_{jj}>0$ and $(i,j)\in\mathcal{E}_p$ (this holds e.g. if agent $i$ is a truth-seeker him/herself since $(i,i)\in\mathcal{E}(W(k))\,\forall k$);
\item if agent $i$ does not satisfy the condition from statement 2, then $x_i(k)\equiv \bar x_i$ for sufficiently large $k$ (the opinion evolution terminates after finite number of iterations);
\item for any $i,j$ either $\bar x_i=\bar x_j$ or $|\bar x_i-\bar x_j|\ge\ve$ (clustering).
\end{enumerate}
\end{theorem}
\begin{proof}
The proof exploits the vector $\xi(k)=(\xi_i(k))_{i=1}^n$ whose elements $\xi_i(k)\dfb |x_i(k)-t|$
stand for the distances from the agents' opinions to the truth.
The sequence $\xi(k)$ appears to be a feasible solution to RAI~\eqref{eq.ineq}, which enables us
to use the result of Theorem~\ref{thm.dynamic-ineq}. Indeed,
\be\label{eq.aux3}
\begin{gathered}
|x_i(k+1)-t|=(1-a_{ii})\left|\sum_{j=1}^n w_{ij}(k)(x_j(k)-t)\right|\\
\leq (1-a_{ii})\sum_{j=1}^n w_{ij}(k)|x_j(k)-t|\quad\forall i\,\forall k.
\end{gathered}
\ee

\textbf{Step 1.} We are going to prove the sufficiency part of statement 2. In view of Proposition~\ref{prop.aux} and Theorem~\ref{thm.dynamic-ineq}, the sequence $\xi(k)$ converges and remains bounded, since $\xi(k)\ge 0$. Therefore, the residual term~\eqref{eq.delta} vanishes, i.e.,
\[
\Delta_i(k)=\sum_{j=1}^n w_{ij}(k)\xi_j(k)-\xi_i(k+1)\xrightarrow[k\to\infty]{}0\quad\forall i.
\]
In view of~\eqref{eq.aux3}, $\Delta_i(k)\geq a_{ii}\sum_jw_{ij}(k)\xi_j(k)\ge a_{ii}w_{ii}(k)\xi_i(k)$. If agent $i$ is a truth-seeker ($a_{ii}>0$), then $\xi_i(k)=|x_i(k)-t|\xrightarrow[k\to\infty]{}0$. If $i$ persistently interacts to a truth-seeker $j$, then $\xi_j(k)$ and $\xi_i(k)$ reach ``consensus''
\[
\lim_{k\to\infty}\xi_i(k)=\lim_{k\to\infty}\xi_j(k)=0.
\]
Therefore, $\lim\limits_{k\to\infty} x_i=t$ for truth-seekers and agents persistently interacting to them.

\textbf{Step 2.} To prove statements 1 and 3, consider the set $I$ of all agents that do not obey the condition of statement 2. For any $i\in I$, one has $a_{ii}=0$. Also, if $i\in I$ and $a_{jj}>0$, then $(i,j)\not\in\mathcal{E}_p$ and hence $w_{ij}(k)=0$ for large $k\ge 0$. An integer $k_0\ge 0$ thus exists such that
\[
x_i(k+1)=\sum_{j\in I}w_{ij}(k)x_j(k),\quad\forall k\ge k_0.
\]
In other words, the subvector $\tilde x(k)=(x_i(k))_{i\in I}$ (after $k_0$ steps) obeys the conventional Hegselmann-Krause model without truth-seekers~\eqref{eq.hk-1}, which terminates in finite time. We have proved statement 3 and, due to statement 1, the existence of limits $\bar x_i$ for all $i$. From the aforementioned properties of the Hegselmann-Krause model~\eqref{eq.hk-1}, we know that
$|\bar x_i-\bar x_j|\in\{0\}\cup[\ve,\infty]$ for any $i,j\in I$.

\textbf{Step 3.} To prove statement 4 and necessity in statement 2, it remains to show that $|\bar x_i-t|\ge\ve$ for any $i\in I$. Suppose, on the contrary, that $|\bar x_i-t|<\ve$. For every truth-seeker $j$, we have
$\lim_{k\to\infty}|x_i(k)-x_j(k)|<\ve$ and thus $w_{ij}(k)>0$ for $k$ being sufficiently large, which contradicts to the assumption $(i,j)\not\in\e_p$.\epf
\end{proof}

Notice that the condition in statement~2 is not easy to check, since $\mathcal{E}_p$ depends on a specific trajectory, which, in turn, nonlinearly depends on the initial condition. To disclose an explicit relation between the terminal opinion profile $\bar x$ and the initial condition $x(0)$
is a non-trivial open problem that remains beyond the scope of this paper.

\subsection{The discrete-time model of bimodal polarization}

Altafini's model of opinion formation, originally proposed in~\cite{Altafini:2012} in its continuous-time form, portrays polarization, or ``bipartite consensus'', of opinions over \emph{structurally balanced} signed graphs~\citep{ShiAltafiniBaras:19}. In this subsection, we consider a discrete-time modification of the Altafini model, examined in~\citep{MengShiCao:16,LiuChenBasar:2017}. This model is similar to the consensus protocol~\eqref{eq.degroot-classic}, allowing, however, negative influence weights
\be\label{eq.altaf}
\begin{gathered}
x(k+1)=A(k)x(k)\in\r^n.
\end{gathered}
\ee
The matrix $A(k)$ satisfies the following assumption.
\begin{assum}
For any $k\ge 0$, the diagonal entries are non-negative $a_{ii}(k)\ge 0$. The non-negative matrix of absolute values $W(k)=(|a_{ij}(k)|)$ is row-stochastic.
\end{assum}

The non-diagonal entries $a_{ij}(k)$ in~\eqref{eq.altaf} may be both positive and negative. Considering the elements  $x_i(k)$ as opinions, the positive value
$a_{ij}(k)>0$ can be treated as trust or friendship between agents $i$ and $j$. In this case, agent $i$ shifts his/her opinion towards the opinion of agent $j$. Similarly, the negative value $a_{ij}(k)<0$ stands for distrust or enmity among the agents: the $i$th agent's opinion is shifted away from the opinion of agent $j$. This rule is motivated by the theory of structural balance in social networks, originating from~\citep{Heider:1944,Heider:1946}.

The central question concerned with the model~\eqref{eq.altaf}, as well as its continuous-time
counterparts~\citep{Altafini:2013,ProCao:2014} is reaching modulus consensus or consensus in absolute value:
\be\label{eq.consen-mod}
\lim_{k\to\infty}|x_1(k)|=\ldots=\lim_{k\to\infty}|x_n(k)|\quad\forall x(0).
\ee
Consensus in absolute value implies that the opinions either reach consensus or polarize: some opinions converge to $M\geq 0$, the other reaching $(-M)$, where the value $M$ depends on the initial condition. It is possible that $M=0$ for all initial conditions, such a situation is considered as degenerate, since the ``consensus'' opinion is independent of the individual opinions of the agents. The following criterion of consensus in absolute value extends the results established in~\citep{MengShiCao:16,LiuChenBasar:2017}.

\begin{theorem}\label{thm.altaf}
Let the sequence $\{W(k)\}$ satisfy Assumptions~\ref{ass.aperiod},\ref{ass.unipos} and~\ref{ass.recipro}. Then the finite limit $\bar x_i=\lim_{k\to\infty}x_i(k)\in\r$ exists for any $i$. If $\g_p$ is strongly connected, then
consensus in absolute value~\eqref{eq.consen-mod} is established and, furthermore,
\begin{itemize}
\item either the system~\eqref{eq.altaf} is globally asymptotically stable, so that  $x(k)\xrightarrow[k\to\infty]{}0\,\forall x(0)$,
\item or the sequence of signed graphs $\g[A(k)]$ is \emph{jointly structurally balanced}~\citep{LiuChenBasar:2017}, that is, a diagonal matrix $D$ with entries $d_{ii}\in\{-1,1\}$ such that $W(k)=DA(k)D$ for $k$ sufficiently large\footnote{In the case of joint structural balance, one has $\sgn a_{ij}(k)=\sgn d_{ii}\sgn d_{jj}$ whenever $a_{ij}\ne 0$. In other words, the relation between agents $i$ and $j$ is positive if $d_{ii},d_{jj}$ have the same sign and negative if the signs are different. The sets of agents with $d_{ii}=1$ and $d_{ii}=-1$ thus constitute two opposing factions in the signed graph $\g[A(k)]$~\citep{Altafini:2013,ProMatvCao:2016}, which condition is referred to as \emph{structural balance}. Opinions in the two factions converge to the opposite values $\bar\xi$ and $-\bar\xi$, where $\bar\xi$ depends on the initial condition; such a property is called ``bipartite consensus''~\cite{Altafini:2013} or bimodal polarization.}.
\end{itemize}
\end{theorem}
\begin{proof}
The columns $\xi(k)=(\xi_i(k))$ of absolute values $\xi_i(k)=|x_i(k)|$ obey RAI~\eqref{eq.ineq} with $w_{ij}(k)=|a_{ij}(k)|$, since
\be
\xi_i(k+1)\le \sum_{j=1}^n|a_{ij}(k)|\xi_j(k)\quad\forall i.
\ee
Thanks to Theorem~\ref{thm.dynamic-ineq}, the finite limits exist $\bar\xi_i=\lim\limits_{k\to\infty}\xi(k)\ge 0$ and
the residuals vanish
\be\label{eq.aux4d}
\Delta_i(k)=\sum_{j=1}^n|a_{ij}(k)|\xi_j(k)-\xi_j(k+1)\xrightarrow[k\to\infty]{}0.
\ee
We are going to show that $x_i(k)$ converges. In the case where $\bar\xi_i=0$, this is obvious.
Assume that $\bar\xi_i\ne 0$ and let $\sigma_i(k)=\sgn x_i(k)$. For any two numbers $c,d\in\r$ such that $c+d\ge 0$ one has $|c|+|d|-(c+d)\ge \min(-2c,0)$. Since
\[
\begin{gathered}
\xi_i(k+1)=x_i(k+1)\sigma_i(k+1)\leq \underbrace{a_{ii}(k)\xi_i(k)\sigma_i(k)\sigma_i(k+1)}_{=c}+\\+\underbrace{\sum_{j\ne 0}|a_{ij}(k)|\xi_j(k)}_{=d},
\end{gathered}
\]
we obtain that $\Delta_i(k)\geq |c|+|d|-(c+d)\geq\min(-2a_{ii}(k)\xi_i(k)\sigma_i(k)\sigma_i(k+1),0)$. According to Assumption~\ref{ass.aperiod}, for any $k$ such that
$\sigma_i(k)\sigma_i(k+1)=-1$ (that is, $x_i$ changes its sign) $\Delta_i(k)\ge 2\eta\xi_i(k)$. By assumption, $\bar\xi_i>0$ and thus for large $k$ one has $\sigma_i(k)\sigma_i(k+1)=1$ in view of~\eqref{eq.aux4d}, in other words, $\sigma_i(k)\equiv\bar\sigma_i$ is constant and therefore $x_i(k)\xrightarrow[k\to\infty]{}\bar x_i=\bar\xi_i\sigma_i$.
If $\g_p$ is strongly connected, then~\eqref{eq.consen-mod} holds due to Theorem~\ref{thm.dynamic-ineq}.

To prove the final statement, we first establish the following relation
\be\label{eq.aux4}
\begin{gathered}
\max(-a_{ij}(k)\sgn \bar x_i\sgn \bar x_j,0)\xrightarrow[k\to\infty]{}0\quad\forall i,j.
\end{gathered}
\ee
We can assume that $\bar x_i,\bar x_j\ne 0$. Since
\[
\begin{gathered}
0\le\xi_i(k+1)=x_i(k+1)\sigma_i(k+1)\le\\ \le\underbrace{a_{ij}(k)\xi_j(k)\sigma_j(k)\sigma_i(k+1)}_{=c}+\underbrace{\sum_{m\ne j}|a_{im}(k)|\xi_m(k)}_{=d},
\end{gathered}
\]
for any $k\ge 0$ one obtains that $\Delta_i(k)\ge 2\min(a_{ij}(k)\xi_j(k)\sigma_j(k)\sigma_i(k+1),0)$. Recalling that $\xi_j(k)\xrightarrow[k\to\infty]{}\bar\xi_j\ne 0$ and $\sigma_i(k+1)\equiv\bar\sigma_i=\sgn\bar x_i$, $\sigma_j(k+1)\equiv\bar\sigma_j=\sgn\bar x_j$ for $k$ being large, one obtains~\eqref{eq.aux4}. Assume now that $\g_p$ is strongly connected and at least one solution $x(k)$ does not vanish as $k\to\infty$. Due to~\eqref{eq.consen-mod}, all components $x_i(k)$ converge to
limits $\bar x_i\ne 0$. Let $d_{ii}\dfb\sgn\bar x_i$. Due to~\eqref{eq.aux4} and Assumption~\ref{ass.unipos}, for large $k$ either $a_{ij}(k)=0$ or $\sgn a_{ij}(k)=d_{ii}d_{jj}$, and thus $(DA(k)D)_{ij}=|a_{ij}(k)|=w_{ij}(k)$ for any $i,j$.\epf
\end{proof}

As it has been discussed, the reciprocity condition from Assumption~\ref{ass.recipro} holds if the graph is 
repeatedly strongly connected, which is a standard condition for consensus in absolute value~\citep{LiuChenBasar:2017}. 
The repeated strong connectivity is necessary and sufficient for \emph{exponential} convergence of the opinions~\citep{LiuChenBasar:2017}.

\subsection{Constrained consensus and common fixed points of paracontractions}

In this subsection, we consider another application of the RAI, related to the problem of \emph{constrained} or ``optimal'' consensus~\citep{Nedic:10,ShiJohansson:2012,CaSrBuCa:14,LinRen:14}, which includes, as a special case, the distributed solution of linear equations~\citep{LiuMorseNedicBasar:14,MouLiuMorse:15,YouSongTempo:16,Wang2019_ARC}.

\begin{figure}[h]
\center
\includegraphics[height=2.5cm]{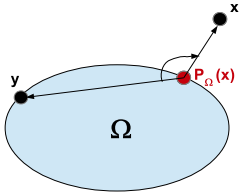}
\caption{The projection onto a closed convex set}\label{fig.proj}
\end{figure}

For any closed convex set $\Omega\subset\r^d$ and $x\in\r^d$ the \emph{projection} operator $P_{\Omega}:\r^d\to\Omega$ maps a point to the closest
element of $\Omega$, i.e., $|x-P_{\Omega}(x)|=\min_{y\in\Omega}|x-y|$. It can be shown that $\measuredangle(y-P_{\Omega}(x),x-P_{\Omega}(x))\ge\pi/2$ (Fig.~\ref{fig.proj}) and
\be\label{eq.proj}
\|x-y\|^2\ge \|x-P_{\Omega}(x)\|^2+\|y-P_{\Omega}(x)\|^2\quad\forall y\in\Omega.
\ee
The distance $d_{\Omega}(x)\dfb\|x-P_{\Omega}(x)\|$ is a convex function.

Consider a group of $n$ agents. Each agent keeps in its memory some constraints, described by a {closed convex} set
$\Xi_i\subseteq\r^d$ (which can be, e.g., a hyperplane in $\r^d$ or the set of minima of some convex function). The agents aim
to find some point $\xi_*\in \Xi_*\dfb \Xi_1\cap\ldots\cap \Xi_n$ that satisfies all the constraints, but do not want to communicate the information about sets $\Xi_i$. Assuming that an agent is able compute the projection $P_i(\xi)=P_{\Xi_i}$ of an arbitrary point $\xi\in\r^d$ onto the set $\Xi_i$, a point belonging to $\Xi_*$ can be computed by one of the
following modifications of the DeGroot iterative procedure~\eqref{eq.degroot-classic}:
\begin{gather}
\xi^i(k+1)=P_i\left[\sum\nolimits_{j=1}^nw_{ij}(k)\xi^j(k)\right]\label{eq.nedic},\\
\xi^i(k+1)=P_i\left[\sum\nolimits_{j=1}^nw_{ij}(k)P_{j}(\xi^j(k))\right]\label{eq.morse},\\
\xi^i(k+1)=w_{ii}(k)P_i(\xi_i(k))+\sum_{j\ne i}w_{ij}(k)\xi^j(k)\label{eq.tempo}.
\end{gather}
The protocol~\eqref{eq.nedic} has been proposed in the influential paper~\cite{Nedic:10}, which deals with distributed optimization problems and then extended (removing some restrictive assumptions) by~\cite{LinRen:14}.
The special cases of protocols~\eqref{eq.morse} and~\eqref{eq.tempo} naturally arise in distributed algorithms, solving linear equations\footnote{In this case, $\Xi_i$ are linear hyperplanes}~\citep{LiuMorseNedicBasar:14,MouLiuMorse:15,YouSongTempo:16}. A randomized version of~\eqref{eq.tempo} was examined by~\cite{ShiJohansson:2012}.
In all of the algorithms, $\xi^i(k)$ stands for an approximation to the desired point $\xi_*$, computed by agent $i$ at step $k$.
In algorithms~\eqref{eq.nedic},\eqref{eq.morse} this approximation always satisfies the constraint of agent $i$
($\xi^i(k)\in\Xi_i$ for $k\ge 1$), whereas~\eqref{eq.tempo} provides this constraint only asymptotically.

We say that \emph{constrained consensus} is established if the sequences $\xi^i(k)$ converge and
\be\label{eq.cons-cons}
\lim_{k\to\infty}\xi^1(k)=\ldots=\lim_{k\to\infty}\xi^n(k)\in\Xi_*.
\ee
It has been recently realized in~\citep{FullmerMorse2018} that the problem of constrained consensus can be considered as a special case of the more general problem of finding a common point for a finite family of \emph{paracontractions}, also known as  $M$-Fejer mappings~\citep{WangRenDuan2019}.
\begin{definition}
A continuous map $M:\r^d\to\r^d$ with the set of fixed points $\mathcal F(M)=\{\zeta:M(\zeta)=\zeta\}$
is a \emph{paracontraction} with respect to some norm $\|\cdot\|$ if
\be\label{eq.para}
\|M(\xi)-M(\xi_0)\|<\|\xi-\xi_0\|\;\;\forall \xi\not\in\mathcal{F}(M),\xi_0\in\mathcal{F}(M).
\ee
\end{definition}
Simple examples of paracontractions are a continuous map without fixed points $\mathcal{F}(M)=\emptyset$ and a contractive mapping ($\|M(\xi)-M(\zeta)\|\le q\|\xi-\zeta\|\,\forall \xi,\zeta\in\r^d$ with $q\in(0,1)$). In the most interesting situations, however, the fixed point is non-unique. The inequality~\eqref{eq.proj} implies that the orthogonal projection $P_{\Omega}$ onto a closed convex set is a paracontraction in the standard Euclidean norm with $\mathcal{F}(P_{\Omega})=P_{\Omega}$. Other examples include, but are not limited to, proximal mappings and gradient descent mappings corresponding to special convex functions, see~\citep{FullmerMorse2018}. Obviously,~\eqref{eq.para} implies that $M$ is non-expansive
\be\label{eq.para+}
\|M(\xi)-M(\xi_0)\|\leq\|\xi-\xi_0\|\quad\forall \xi\in\r^d,\xi_0\in\mathcal{F}(M).
\ee
Notice that the requirement of constrained consensus~\eqref{eq.cons-cons} can be reformulated as follows: the distributed algorithm converges to a common \emph{fixed point} of the paracontractive projection operators $P_i=P_{\Xi_i}$. A natural question arises whether the algorithms~\eqref{eq.nedic}--\eqref{eq.tempo} (under proper assumptions on $W(k)$) are capable of computing a common fixed point of a general family of paracontractions $P_i$, that is, an element of $\Xi_*=\bigcap\mathcal{F}(P_i)$? For the algorithm~\eqref{eq.nedic},
the affirmative answer was given in~\citep{FullmerMorse2018}, assuming that
the graph is repeatedly strongly connected and $\|\cdot\|=\|\cdot\|_p$ with some $1\le p\le\infty$.
We can here extend this result to all algorithms~\eqref{eq.nedic}--\eqref{eq.tempo} and an arbitrary norm on $\mathbb{R}^d$.

Analysis of the algorithms~\eqref{eq.nedic}--\eqref{eq.tempo} relies on a technical lemma,
which holds for repeatedly strongly (and even \emph{quasi-strongly}) connected graphs.
This lemma establishes robustness of the iterative averaging procedure~\eqref{eq.degroot-classic}
against asymptotically vanishing disturbances.

\begin{lemma}~\citep{LinRen:14}\footnote{Formally, Lemma~9 in~\citep{LinRen:14} adopts some stronger requirements on the
matrix $W(k)$ than Lemma~\ref{lem.aux5}, however, its proof employs only the exponential convergence of backward matrix products
$W(k)\ldots W(0)$, which can be established~\citep{Blondel:05} under the assumptions of  Lemma~\ref{lem.repeated-qs}.
An alternative proof can be given by retracing the arguments from~\citep{ShiJohansson:13}, concerned with robustness of
continuous-time consensus algorithms.}.\label{lem.aux5}
Let the matrices $W(k)$ satisfy the conditions of Lemma~\ref{lem.repeated-qs} and consider such sequences of vectors $\xi^1(k),\ldots,\xi^n(k)\in\r^d$, $k\geq 0$, that
\be\label{eq.conse-weak}
\xi^i(k+1)=\sum_{j=1}^nw_{ij}(k)\xi^j(k)+e^i(k)\,\forall i,\; e^i(k)\xrightarrow[k\to\infty]{}0.
\ee
Then the sequences $\xi^i(k)$ asymptotically synchronize, that is, $\lim_{k\to\infty}\|\xi^i(k)-\xi^j(k)\|=0\,\forall i,j$.
\end{lemma}

Note that Lemma~\ref{lem.aux5} does not guarantee the convergence of $\xi^i(k)$;
the latter property requires stronger assumptions on the ``disturbances'' $e^i(k)$.
To prove convergence of the algorithms, we also use the following proposition.
\begin{proposition}\label{prop.aux5}
Let $M$ be a paracontraction in some norm $\|\cdot\|$ and $\xi^0$ be its fixed point. Denote
$d(\xi)\dfb \|\xi-\xi^0\|-\|M\xi-\xi^0\|\geq 0$ and consider a \emph{bounded} sequence of vectors $\xi(k)$ such
that $d(\xi(k))\xrightarrow[k\to\infty]{}0$. Then,
\[
\|M\xi(k)-\xi(k)\|\xrightarrow[k\to\infty]{}0.
\]
\end{proposition}
\begin{proof}
Assume, on the contrary, that $
\|M\xi(k_r)-\xi(k_r)\|\ge\ve$, $\forall r=1,2,\ldots,
$ for a number $\ve>0$ and a sequence $k_r\to\infty$. Passing
to a subsequence, one can assume without loss of generality that the vectors $\xi(k_r)$ converge to
a limit $\xi_*\in\r^d$. Recalling that $M$ is continuous, one has $\|M\xi_*-\xi_*\|\ge\ve$ while $d(\xi_*)=0$.
One arrives at a contradiction with~\eqref{eq.para}, since $\xi_*\not\in\mathcal{F}(M)$
whereas $\|M\xi_*-\xi^0\|=\|\xi_*-\xi^0\|$.\epf
\end{proof}

We now formulate the main result of this subsection.
\begin{theorem}\label{thm.para}
Suppose that maps $P_i:\r^d\to\r^d$ are paracontractions with respect to some common norm $\|\cdot\|$ that have at least one common fixed point $\Xi_*=\bigcap_{i=1}^n\mathcal{F}(P_i)\ne\emptyset$. Let the assumptions of Corollary~\ref{cor.repeated}
hold. Then each of the algorithms~\eqref{eq.nedic}--\eqref{eq.tempo} finds a common fixed point of $\{P_i\}$, that is,~\eqref{eq.cons-cons} holds.
\end{theorem}
\begin{proof}
We first introduce some auxiliary notation. Fix an arbitrary point $\xi^0\in\Xi_*$. For this point, let
$\delta_i(\xi)\dfb \|\xi-\xi^0\|-\|P_i\xi-\xi^0\|\geq 0$. Let $\zeta^i(k)=\sum_{j=1}^nw_{ij}(k)\xi^j(k)$.
{The central idea of the proof is to explore the properties of the vectors $x(k)=(x_i(k))_{i=1}^n$ whose components
$x_i(k)\dfb \|\xi^i(k)-\xi^0\|$ stand for the distances of the agents' ``opinions'' $\xi^i(k)$ to the chosen fixed point.
It will be shown (Step 1) that these vectors satisfy the  RAI~\eqref{eq.ineq}, which enable us to use the techniques of Theorem~\ref{thm.dynamic-ineq}.}

\textbf{Step 1.} We are going to show first that the sequence $x(k)$ is a feasible solution to the RAI~\eqref{eq.ineq}. In the case of algorithm~\eqref{eq.nedic},
one has
\be\label{eq.nedic-i}
\begin{gathered}
x_i(k+1)=\|P_i(\zeta^i(k))-\xi^0\|\overset{\eqref{eq.para+}}{\leq}\|\zeta^i(k)-\xi^0\|=\\
=\left\|\sum_jw_{ij}(k)(\xi^j(k)-\xi^0)\right\|\leq\\
\leq \sum_jw_{ij}(k)\|\xi^j(k)-\xi^0\|=\sum_jw_{ij}(k)x_j(k).
\end{gathered}
\ee
The case of~\eqref{eq.morse} is considered similarly. Denoting $\bar\zeta^i(k)\dfb\sum_{j=1}^nw_{ij}(k)P_j(\xi^j(k))$,
\be\label{eq.morse_i}
\begin{gathered}
x_i(k+1)=\|P_i(\bar\zeta^i(k))-\xi^0\|\overset{\eqref{eq.para+}}{\leq}\|\bar\zeta^i(k)-\xi^0\|=\\
=\left\|\sum_jw_{ij}(k)(P_j(\xi^j(k))-\xi^0)\right\|\leq \\ \leq \sum_jw_{ij}(k)\|P_j(\xi^j(k))-\xi^0\|\overset{\eqref{eq.para+}}{\leq}
\\\leq
\sum_jw_{ij}(k)\|\xi^j(k)-\xi^0\|=\sum_jw_{ij}(k)x_j(k).
\end{gathered}
\ee
In the case of algorithm~\eqref{eq.tempo}, one has
\be\label{eq.tempo-i}
\begin{gathered}
x_i(k+1)=\left\|w_{ii}(k)(P_i(\xi^i(k))-\xi^0))+\sum_{j\ne i}w_{ij}(k)(\xi^j(k)-\xi^0)\right\|\\
\leq w_{ii}(k)\|P_i(\xi^i(k))-\xi^0)\|+\sum_{j\ne i}w_{ij}(k)\|\xi^j(k)-\xi^0\|\overset{\eqref{eq.para+}}{\leq}\\
\leq w_{ii}(k)\|\xi^i(k)-\xi^0\|+\sum_{j\ne i}w_{ij}(k)x_j(k)=\sum_{j=1}^nw_{ij}(k)x_j(k).
\end{gathered}
\ee
Using Corollary~\ref{cor.repeated}, one shows that the RAI establishes consensus, that is,
$x_i(k)\xrightarrow[k\to\infty]{} c\ge 0\,\forall i$ and $\Delta_i(k)\xrightarrow[k\to\infty]{} 0$.

\textbf{Step 2.} Using~\eqref{eq.resid0}, we are now going to prove~\eqref{eq.conse-weak}, which can be also written as
\be\label{eq.aux5+}
\|\xi^i(k+1)-\zeta^i(k)\|\xrightarrow[k\to\infty]{}0\quad\forall i.
\ee
Also, it will be shown that
\be\label{eq.aux5++}
\|P_i(\xi^i(k))-\xi^i(k)\|\xrightarrow[k\to\infty]{}0\quad\forall i.
\ee

Notice first that the vectors $\xi^i(k)$ and $\zeta^i(k)$ are bounded, since their distances $x_i(k)$ to $\xi^0$ are bounded.
For algorithm~\eqref{eq.nedic}, the inequality \eqref{eq.nedic-i} imply that
\be\label{eq.aux6-1}
\Delta_i(k)\ge \|\zeta^i(k)-\xi^0\|-\|P_i(\zeta^i(k))-\xi^0\|=\delta_i(\zeta^i(k)).
\ee
Applying Proposition~\ref{prop.aux5} to $M=P_i$,~\eqref{eq.resid0} leads
to $P_i(\zeta^i(k))-\zeta^i(k)\xrightarrow[k\to\infty]{}0$, which is equivalent to~\eqref{eq.aux5+} ($\xi_i(k+1)=P_i(\zeta^i(k))$). To derive~\eqref{eq.aux5++}, note that
\[
\begin{gathered}
\|P_i(\xi^i(k+1))-\xi^i(k+1)\|\leq \underbrace{\|P_i(\xi^i(k+1))-P_i(\zeta^i(k))\|}_{\leq\|\xi^i(k+1)-\zeta^i(k)\|}+\\+\|P_i(\zeta^i(k))-\zeta^i(k)\|
+\|\zeta^i(k)-\xi^i(k+1)\|\xrightarrow[k\to\infty]{}0.
\end{gathered}
\]
In the case of algorithm~\eqref{eq.tempo}, we notice that~\eqref{eq.tempo-i} yields
\be\label{eq.aux6-2}
\begin{gathered}
\Delta_i(k)\ge w_{ii}(k)(\|\xi^i(k)-\xi^0\|-\|P_i(\xi^i(k))-\xi^0\|)\geq\\\geq\eta\delta_i(\xi^i(k)).
\end{gathered}
\ee
Here $\eta>0$ is the constant from Assumption~\ref{ass.aperiod}. Applying Proposition~\ref{prop.aux5}, one
proves~\eqref{eq.aux5++}, entailing also~\eqref{eq.aux5+} since $\xi^i(k+1)=\zeta^i(k)+w_{ii}(k)(P_i(\xi^i(k))-\xi^i(k))$.

The proof in the case of algorithm~\eqref{eq.morse} combines the two aforementioned estimates. First,~\eqref{eq.morse_i}
yields in
\be\label{eq.aux6-3}
\Delta_i(k)\ge \|\bar\zeta^i(k)-\xi^0\|-\|P_i(\bar\zeta^i(k))-\xi^0\|=\delta_i(\bar\zeta^i(k)),
\ee
which  is similar to~\eqref{eq.aux6-1} and entails that
\[
\|\xi_i(k+1)-\bar\zeta_i(k)\|=\|P_i(\bar\zeta^i(k))-\bar\zeta^i(k)\|\xrightarrow[k\to\infty]{}0.
\]
Second,~\eqref{eq.morse_i} also implies~\eqref{eq.aux6-2}, which in turn implies~\eqref{eq.aux5++}. Thus $\|\bar\zeta^i(k)-\zeta^i(k)\|\xrightarrow[k\to\infty]{}0$. This proves~\eqref{eq.aux5+} since
$
\|\xi_i(k+1)-\bar\zeta^i(k)\|\leq \|\xi_i(k+1)-\bar\zeta^i(k)\|+\|\bar\zeta^i(k)-\zeta^i(k)\|.
$

Since~\eqref{eq.aux5+} is equivalent to~\eqref{eq.conse-weak}, Lemma~\ref{lem.aux5} ensures synchronization property $\|\xi^i(k)-\xi^j(k)\|\xrightarrow[k\to\infty]{}0$ for all $i,j$.

\textbf{Step 3.} Recalling that the vectors $\xi^i(k)$ are bounded, there exists a sequence $k_r\to\infty$ such that $\xi^1(k_r)\xrightarrow[r\to\infty]{}\xi_*\in\r^d$. In view of~\eqref{eq.conse-weak}, we have $\xi^i(k_r)\xrightarrow[r\to\infty]{}\xi_*$ for every $i$. The property~\eqref{eq.aux5++} implies that
$P_i(\xi_*)=\xi_*$, and thus $\xi_*\in\Xi_*$. It remains to show that $\xi^i(k)\xrightarrow[k\to\infty]{}\xi_*$.

To prove this, notice that at Step 1 we have not specified the choice of $\xi^0$, which can be an arbitrary point in $\Xi_*$.
We have proved that for \emph{any} such point the distances $x_i(k)=\|\xi^i(k)-\xi^0\|$ converge to some consensus value $c$,
depending on $\xi^0$ and the initial conditions. In particular, substituting $\xi^0=\xi_*$, we obtain that the limits exist
\[
x_i(k)=\lim_{k\to\infty}\|\xi^i(k)-\xi_*\|=c_*\quad\forall i.
\]
Recalling that $x_i(k_r)\xrightarrow[r\to\infty]{}0$, one has $c_*=0$, which shows that the limits in~\eqref{eq.cons-cons}
exist and are equal to $\xi_*\in\Xi_*$. \epf
\end{proof}

\begin{remark}  Since Lemma~\ref{lem.aux5} retains its validity~\citep{LinRen:14} in the case of bounded communication delays,
Theorem~\ref{thm.robust} allows to extend the result of Theorem~\ref{thm.para} to networks with delayed communication.
The corresponding extension is straightforward and is thus omitted.
\end{remark} 
\section{Proofs of the technical results}\label{sec.proofs}
In this section, we prove Theorems~\ref{thm.static-ineq}--\ref{thm.new}. We start with some auxiliary constructions and technical lemmas.

\subsection{Preliminary results}

We start with the following simple proposition.
\begin{proposition}\label{prop.product}
For any sequence $a_1,\ldots,a_m\in [0,1-\eta]$, where $\eta\in (0,1)$, the following inequality holds
\be\label{eq.pi}
\pi(a_1,\ldots,a_m)\dfb\prod_{i=1}^m(1-a_i)\geq \exp\left(-\eta\sum_{i=1}^ma_i\right).
\ee
\end{proposition}
\begin{proof}
Since $\pi\dfb\pi(a_1,\ldots,a_m)>0$, one has
\[
\begin{gathered}
-\log\pi=\sum_{i=1}^m\log\left(\frac{1}{1-a_i}\right)=\sum_{i=1}^m\log\left(1+\frac{a_i}{1-a_i}\right)\overset{(+)}{\leq}\\
\leq \sum_{i=1}^m\frac{a_i}{1-a_i}
\leq\eta\sum_{i=1}^ma_i.
\end{gathered}
\]
The inequality denoted $(+)$ uses the well-known fact that $\log(1+a)\le a$ for any $a\ge 0$.
\end{proof}

We now derive two important estimates for the solutions of the RAI~\eqref{eq.ineq}. For a non-empty set $I\subseteq [1:n]$, denote
\[
M_I(k)\dfb\max_{i\in I}x_i(k),\quad M(k)\dfb \max_ix_i(k)=M_{[1:n]}(k).
\]
Along with the number of arcs $a_{I,J}(k_0:k_1)$ between two non-empty subsets of agents $J,I$, consider the total flow from $J$ to $I$, which is defined as follows
\ben
\w_{I,J}(k_0:k_1)\dfb\sum_{k=k_0}^{k_1}\sum_{i\in I,j\in J}w_{ij}(k).
\een
Let $\w_{I,J}(k)\dfb \w_{I,J}(k:k),\,\w_{i,J}(k_0:k_1)\dfb \w_{\{i\},J}(k_0:k_1)$.
In some sense, the flow $\w_{I,J}(k_0:k_1)$ measures the influence of agents from $J=I^c$ on agents from $I$ during the time interval $[k_0,k_1]$. In particular, if $\w_{I,J}(k_0:k_1)=0$, then the group of agents $I$ evolves independently of $J$ on this interval, so that $M_I(k)$ is non-decreasing (Proposition~\ref{prop.bound}). It can be expected that if $\w_{I,J}(k_0:k_1)$ is small, the maximal increase $M_I(k_1+1)-M_I(k_0)$ can be positive yet admits a small upper estimate. This fact is established by the following simple lemma.

\begin{lemma}\label{lem.maximum}
Assume that Assumption~\ref{ass.aperiod} holds.
For every instants $k_0\ge 0$, $k_0'\ge k_0$, $k_1\ge k_0'$ and every cut $(I,J)$ with $I,J\ne\emptyset$
the solution $x(\cdot)$ obeys the inequality
\be\label{eq.est1}
\begin{gathered}
M_I(k_1+1)\leq \vartheta M_I(k_0')+(1-\vartheta) M(k_0),\\
\vartheta\dfb \exp(-\eta\w_{I,J}(k_0':k_1)),
\end{gathered}
\ee
\end{lemma}
\begin{proof}
We know (Proposition~\ref{prop.bound}) that $x_i(k)\leq M(k)\le M(k_0)\,\forall k\ge k_0\forall i$. Denoting $\rho(k)\dfb\max_{i\in I}\w_{i,J}(k)$,
for each $k\ge k_0$, one obtains the inequality
\be\label{eq.aux-aux}
\begin{gathered}
x_i(k+1)\leq \sum_{j\in I}w_{ij}(k)\underbrace{x_j(k)}_{\leq M_I(k)}+\sum_{j\in J}w_{ij}(k)\underbrace{x_j(k)}_{\leq M(k_0)}=\\=
(1-\w_{i,J}(k))M_I(k)+\w_{i,J}(k)M(k_0)=\\
=M(k_0)-(1-\w_{i,J}(k))(M(k_0)-M_I(k))\leq\\
\leq M(k_0)-(1-\rho(k))(M(k_0)-M_I(k))\quad\forall i\in I.
\end{gathered}
\ee
Therefore, $M(k_0)-M_I(k+1)\geq \left(1-\rho(k)\right)(M(k_0)-M_I(k))\,\forall k\ge k_0$ and thus
\[
\begin{gathered}
M(k_0)-M_I(k_1+1)\geq (M(k_0)-M_I(k_0'))\prod_{k=k_0'}^{k_1}\left(1-\rho(k)\right).
\end{gathered}
\]
By noticing that $\w_{i,J}(k)\leq 1-w_{ii}(k)\leq 1-\eta$ and also $\w_{i,J}(k)\leq \w_{I,J}(k)$ for each $i\in I$, Proposition~\ref{prop.product} yields in
\[
\begin{gathered}
M(k_0)-M_I(k_1+1)\geq (M(k_0)-M_I(k_0'))e^{-\eta\sum\limits_{k=k_0'}^{k_1}\rho(k)}\geq\\
\geq (M(k_0)-M_I(k_0'))e^{-\eta\sum\limits_{k=k_0'}^{k_1}\w_{I,J}(k)}=(M(k_0)-M_I(k_0'))\theta,
\end{gathered}
\]
which inequality is equivalent to~\eqref{eq.est1}.\epf
\end{proof}

Lemma~\ref{lem.maximum} can be extended to delayed RAI~\eqref{eq.ineq-d}, although the estimate is more conservative.
\begin{lemma}\label{lem.maximum-d}
Let Assumption~\ref{ass.aperiod} hold. For a solution of the RAI~\eqref{eq.ineq-d}, denote
\[
\begin{gathered}
\bar M(k)\dfb \max\{M(k),M(k-1),\ldots,M(k-d_*)\},\\
\bar M_I(k)\dfb \max\{M_I(k),\ldots,M_I(k-d_*)\}.
\end{gathered}
\]
For every instants $k_0\geq 0$, $k_0'\geq k_0$ and $k_1\geq k_0$ and every cut $(I,J)$ with $I,J\ne\emptyset$ one has
\be\label{eq.est1-d}
\begin{gathered}
\bar M_I(k_1+1)\leq \bar\vartheta \bar M_I(k_0')+(1-\bar\vartheta) \bar M(k_0),\\
\bar\vartheta\dfb \eta^{d^*}\exp(-\eta^{d_*+1}\w_{I,J}(k_0':k_1)).
\end{gathered}
\ee
\end{lemma}
\begin{proof}

Using~\eqref{eq.ineq-d}, it is easy to show that the sequence $\bar M(k)$ is non-increasing.
Denoting $\rho(k)\dfb\min_{i\in I}\w_{i,J}(k)$,
for each $k\ge k_0$ and $i\in I$, the following inequality is derived from~\eqref{eq.ineq-d} similar to~\eqref{eq.aux-aux}
\be\label{eq.aux-aux1}
\begin{gathered}
x_i(k+1)\leq (1-\w_{i,J}(k))\bar M_I(k_0)+\w_{i,J}(k)\bar M(k_0)\\
\leq \bar M(k_0)-\w_{i,J}(k)(\bar M(k_0)-\bar M_I(k))\leq\\
\leq \bar M(k_0)-\rho(k)(\bar M(k_0)-\bar M_I(k)).
\end{gathered}
\ee
Besides this, notice that for any $k'\ge k_0$ and any $i$
\ben
\begin{aligned}
x_i(k'+1)\leq w_{ii}(k')x_i(k')+(1-w_{ii}(k'))\bar M(k_0)=\\
=\bar M(k_0)-w_{ii}(k')[\bar M(k_0)-x_i(k')]\leq\\
\leq\bar M(k_0)-\eta(\bar M(k_0)-x_i(k')),
\end{aligned}
\een
entailing the estimate
\be\label{eq.aux-aux2}
\bar M(k_0)-x_i(k'+s)\geq \eta^{s}(\bar M(k_0)-x_i(k'))\,\forall s\ge 1.
\ee
Combining~\eqref{eq.aux-aux2} with~\eqref{eq.aux-aux1}, one proves that for $i\in I$ and $k\ge k_0$ the inequality holds as follows
\be\label{eq.aux-aux2+}
\begin{gathered}
x_i(k+s)\leq \bar M(k_0)-\eta^{d_*}\rho(k)(\bar M(k_0)-\bar M_I(k))\\
\forall s=1,\ldots,d_*,
\end{gathered}
\ee
in particular, $\bar M(k_0)-\bar M_I(k+d_*)\geq (1-\eta^{d_*}\rho(k))(\bar M(k_0)-\bar M_I(k))$.
Denote now $m\dfb\lfloor(k_1-k_0')/d_*\rfloor$, so that $k_0'+md_*\leq k_1<k_0'+(m+1)d_*$. Then,
\be\label{eq.aux-aux2++}
\begin{gathered}
\bar M(k_0)-\bar M_I(k_0'+md_*)\geq\\
\geq
(\bar M(k_0)-\bar M_I(k_0'))\prod_{j=0}^{m-1}(1-\eta^{d_*}\rho(k_0'+jd_*))\geq
\\
\geq e^{-\eta^{(d_*+1)}\sum\limits_{j=0}^{m-1}\rho(k_0'+jd_*)}(\bar M(k_0)-\bar M_I(k_0'))\geq\\
\geq e^{-\eta^{(d_*+1)}\w_{I,J}(k_0':k_1)}(\bar M(k_0)-\bar M_I(k_0'))=\\=
\bar\vartheta\eta^{-d_*}(\bar M(k_0)-\bar M_I(k_0')),
\end{gathered}
\ee
or, equivalently, $\bar M(k_0)-x_i(k_0'+md_*)\geq \bar\vartheta\eta^{-d_*}(\bar M(k_0)-\bar M_I(k))$ for each $i\in I$.
Substituting $k'=k_0'+md_*$ and $s=1,\ldots,k_1+1-k'\le d_*$ into~\eqref{eq.aux-aux2+}, we have
\be\label{eq.aux-aux2+++}
\begin{gathered}
\bar M(k_0)-x_i(k)\geq \bar\vartheta(\bar M(k_0)-\bar M_I(k_0'))\,\forall i\in I,
\end{gathered}
\ee
whenever $k\in [(k'+1):(k_1+1)]$. In view of~\eqref{eq.aux-aux2++},~\eqref{eq.aux-aux2+++} holds also for all $k\in [(k_1-d_*):k_1]$,
entailing thus~\eqref{eq.est1-d}.
\end{proof}

Combining the results of Lemmas~\ref{lem.maximum},\ref{lem.maximum-d} with Assumptions~\ref{ass.unipos} and~\ref{ass.recipro}, we are now able to derive the following corollary, lying in the heart  of the proofs of Theorems~\ref{thm.dynamic-ineq} and~\ref{thm.robust}. Informally, these corollary mean that for a strongly connected persistent graph, the maximum of
$\bar M(k)$ (or $M(k)$ in the undelayed case) will eventually decrease until the opinions become unanimous.
We formulate the corresponding result for RAI~\eqref{eq.ineq-d}, which includes the undelayed RAI~\eqref{eq.ineq-d} as a special case ($d_*=0$).

\begin{corollary}\label{cor.maximum}
Suppose that Assumptions~\ref{ass.aperiod},\ref{ass.unipos} and~\ref{ass.recipro} hold and $\g_p$ is strongly connected. Assume that for some $i\in [1:n]$ and some $k_0\ge 0$, a feasible solution of RAI~\eqref{eq.ineq-d} satisfies the condition $\bar M(k_0)-x_i(k_0)\ge\ve>0$. Then there exist $k_*\ge k_0$ and constant $\varrho\in (0,1)$, independent of $\ve,k_0$, such that inequality holds as follows
\be\label{eq.aux7}
\bar M(k_*+1)\le \bar M(k_0)-\varrho\ve.
\ee
\end{corollary}
\begin{proof}
\textbf{Step~1.} First, we prove the following statement: \emph{Consider a cut $(I,J)$ with $I,J\ne\emptyset$ and two instants $k_0\ge 0$, $k_0'\geq k_0$ such that $\bar M(k_0)-x_i(k_0')\ge\ve>0$ for any $i\in I$. Then there exist $k_1\geq k_0'$ and $j_0\in J$ such that
\be\label{eq.aux7d}
x_i(k_1+1)\le \bar M(k_0)-\varrho_0\ve\quad\forall i\in I_1=I\cup\{j_0\},
\ee
where the constant $\varrho_0$ is determined by $\eta$ from Assumptions~\ref{ass.aperiod},~\ref{ass.unipos} and $\M,T$ from Assumption~\ref{ass.recipro}.}

Notice first that, due to~\eqref{eq.aux-aux2},
\begin{equation}\label{eq.aux-aux-aux}
\begin{gathered}
\bar M(k_0)-x_i(k_0'+s)\geq\eta^s (\bar M(k_0)-x_i(k_0'))\geq\eta^{d_*}\ve\\
\forall i\in I\quad\forall s=0,\ldots,d_*.
\end{gathered}
\end{equation}
Thus, for $k_0''\dfb k_0'+d_*$ one has $\bar M_I(k_0'')\leq \bar M(k_0)-\eta^{d_*}\ve$.

Since $\g_p$ is strongly connected, $a_{J,I}(k_0':\infty)=\infty$. Let $k_1\ge k_0''$ be the first instant such that $a_{J,I}(k_0'':k_1)>0$. In view of Assumption~\ref{ass.recipro} and~\eqref{eq.gen-balance}, either $k_1-k_0''<T$ and thus
$a_{I,J}(k_0'':k_1)\leq |I||J|T\leq (|I|+|J|)^2T/4=n^2T/4$ or $a_{I,J}(k_0'':(k_1-T))<\M$. In both cases, we obtain that $\w_{I,J}(k_0'':k_1)\leq a_{I,J}(k_0'':k_1)<\M_0\dfb \max(\M,n^2T/4)$.

In view of the estimates on $\bar M_I(k_0'')$ and $\w_{I,J}(k_0'':k_1)$,
\begin{equation}\label{eq.aux-aux-aux1+}
\begin{gathered}
\bar M_I(k)\leq \bar M(k_0)-\bar\theta'\eta^{d_*}\ve\;\;\forall k\in [k_0'':(k_1+1)]\\
\bar\vartheta'\dfb \eta^{d^*}\exp(-\eta^{d_*+1}\M_0).
\end{gathered}
\end{equation}
Inequality~\eqref{eq.aux-aux-aux1+} is obvious if $k=k_0''$ (since $\bar\theta'<1$ and $\bar M_I(k_0'')\leq \bar M(k_0)-\eta^{d_*}\ve$), otherwise it follows from Lemma~\ref{lem.maximum-d} applied to the time interval $[k_0'':(k-1)$.

By definition of $k_1$, there exists an arc $(i_0,j_0)$ in $\g[W(k_1)]$ connecting $i_0\in I$ to some $j_0\in J$.  Denoting $d_0\dfb d_{j_0i_0}(k_1)$ and $w_0\dfb w_{j_0i_0}(k_1)\ge \eta$, one arrives at
\be
\begin{gathered}
x_{j_0}(k_1+1)\leq w_0\underbrace{x_{i_0}(k_1-d_0)}_{\leq M_I(k_1)}+(1-w_0(k_1))\bar M(k_0)\leq\\
\leq\bar M(k_0)-\varrho_0\ve,\\
\varrho_0\dfb\eta^{d_*+1}\bar\theta'=\eta^{2d_*+1}e^{-\eta^{d_*}\M_0}.
\end{gathered}
\ee
Along with~\eqref{eq.aux-aux-aux1+}, the latter inequality implies~\eqref{eq.aux7d}.

\textbf{Step~2.} Corollary~\ref{cor.maximum} is now proved by the following inductive procedure.
Applying the statement proved at Step~1 to $k_0'=k_0$ and $I_0=\{i\}$, one proves the existence of $k_1\geq k_0$ and the set $I_1\supsetneq I_0$ of cardinality $|I_1|=2$ such that~\eqref{eq.aux7d} holds. Applying the same statement to $I=I_1$, $k_0'=k_1+1$ one shows the existence of an instant $k_2\geq k_1+1$ and a set $I_2\supsetneq I_1$ of cardinality $|I_2|=3$ such that
\be\label{eq.aux7+}
x_i(k_2+1)\le M(k_0)-\varrho_0^2\ve\quad\forall i\in I_2,
\ee
and so on. Iterating this argument $n-1$ times, one finally arrives at the existence of $k_n\geq k_{n-1}+1>k_0$ such that
\be\label{eq.aux7+}
x_i(k_n+1)\le M(k_0)-\varrho_0^{n-1}\ve\quad\forall i\in I_n=[1:n],
\ee
entailing that $\bar M(k_n+d_*)\leq \bar M(k_0)-\eta^{d*}\varrho_0^{n-1}\ve$ due to~\eqref{eq.aux-aux2}.
Hence,~\eqref{eq.aux7} holds with $\varrho=\eta^{d*}\varrho_0^{n-1}$.\epf
\end{proof}

\subsection{Proof of Lemma~\ref{lem.persist1}}

To prove the first statement, assume that the RAI~\eqref{eq.ineq} is convergent yet $\g_p$ contains a strong component with outcoming arcs. Then a \emph{source} component exists that has an outcoming arc but no incoming ones.~\footnote{In other words, if the acyclic \emph{condensed graph}~\cite{HararyBook:1965} has at least one arc, it should have a source node with at least one arc coming from it.} Denote the set of its nodes by $I\subsetneq [1:n]$ and let $J\dfb I^c$. Since $(j,i)\not\in\e_p$ for any $i\in I,j\in J$, one has $\w_{I,J}(0:\infty)<\infty$. Denote $a_k=\max_{i\in I}\w_{i,J}(k)$.
In view of Assumption~\ref{ass.aperiod}, $\w_{i,J}(k)\le 1-w_{ii}(k)\le 1-\eta$. Defining $\zeta(0)\dfb 1$ and
\[
\zeta(m)\dfb\prod_{k=0}^{m-1}(1-a_k),\quad 1\leq m\leq\infty.
\]
the sequence $\zeta(m)$ is decreasing and converges to a limit $\zeta(\infty)$; Proposition~\ref{prop.product} entails that that $\zeta(\infty)>0$ since $\sum_ka_k<\infty$. Notice also that
\be\label{eq.aux8}
\begin{aligned}
\zeta(k+1)=(1-a_k)\zeta(k)&\leq\zeta(k)-\sum_{j\in J}w_{ij}(k)\zeta(k)=\\
&=\sum_{j\in I}w_{ij}(k)\zeta(k)\quad\forall i\in I.
\end{aligned}
\ee
By assumption, there exists a persistent arc $(i_0,j_0)\in\e_p$, where $i_0\in I$ and $j_0\in J$, therefore, $w_{j_0i_0}(k)\geq\eta$ for an infinite sequence of instants $k=k_1,k_2,\ldots$. We will now construct a solution to~\eqref{eq.ineq} that does not converge. Let
\[
x_i(k)=
\begin{cases}
\zeta(k),\quad &i\in I,\\
0,\quad &i\in J\setminus\{j_0\},\\
0,\quad &(i=j_0)\land (k\ne k_s+1\forall s),\\
\frac{1+(-1)^s}{2}\eta\zeta(\infty),\quad &(i=j_0)\land (k=k_s+1).
\end{cases}
\]
Obviously, $x_{j_0}(k_s+1)$ does not converge as $s\to\infty$, so $x(k)$ fails to have a limit as $k\to\infty$. We are going to prove that $x(k)$ is a solution to the RAI, that is,
\be\label{eq.aux8+}
x_i(k+1)\leq\sum_{j=1}^nw_{ij}(k)x_j(k).
\ee
For $i\in I$, the latter inequality follows from~\eqref{eq.aux8} since
\[
\begin{gathered}
x_i(k+1)\overset{\eqref{eq.aux8}}{\leq} \sum_{j\in I}w_{ij}(k)x_j(k)\leq \sum_{j\in I}w_{ij}(k)x_j(k)+\\
+\sum_{j\in J}w_{ij}(k)\underbrace{x_j(k)}_{\geq 0}=\sum_{j=1}^nw_{ij}(k)x_j(k).
\end{gathered}
\]
Notice also that $0\leq x_{j_0}(k+1)\leq w_{j_0i_0}(k)x_{i_0}(k)$ for any $k$. For $k\ne k_s$, this is obvious since $x_{i_0}(k)\ge 0=x_{j_0}(k+1)$. For $k=k_s$, one has $w_{j_0i_0}(k)\geq\eta$ and $x_{i_0}(k)>\zeta(\infty)$. Since all components $x_i(k)$ are non-negative,~\eqref{eq.aux8+} holds also for $i=j_0$. For $i\in J\setminus\{j_0\}$,~\eqref{eq.aux8+} is obvious since $x_j(k)\ge 0\,\forall j$.

The second statement is immediate from Lemma~\ref{lem.persist}. If consensus is established by the RAI~\eqref{eq.ineq}, it is automatically established by the iterative averaging procedure~\eqref{eq.degroot-classic}, and hence $\g_p$ is quasi-strongly connected. Since all strong components of $\g_p$ are isolated, $\g_p$ has to be strong.

\begin{remark}\label{rem.delay}
The construction of the oscillatory solution works also for delayed RAI~\eqref{eq.ineq-d}, adding the initial conditions: $x_i(-1)=\ldots=x_i(-d_*)=\zeta(0)=1$ for $i\in I$, $x_i(-1)=\ldots=x_i(-d_*)=0$ for $i\not\in I$.
In this case, $x_i(k-d)\ge \xi(k)>\xi(\infty)$ for any $k\ge 0$, $0\le d\le d_*$, $i\in I$.
\end{remark}
\begin{remark}\label{rem.aperiod}
In the proof of Lemma~\ref{lem.persist1}, we in fact used a relaxed form of Assumption~\ref{ass.aperiod}: $w_{ii}(k)\ge\eta$ for each node of the graph, belonging to a \emph{source} component. This will be used in the proof of Theorem~\ref{thm.static-ineq}.
\end{remark}

\subsection{Proof of Theorem~\ref{thm.dynamic-ineq}}

We first notice that Assumptions~\ref{ass.unipos} and~\ref{ass.recipro} imply that the strong components of $\g_p$ are isolated. Indeed,~\eqref{eq.gen-balance} entails that if $a_{I,J}(0:\infty)=\infty$, then $a_{J,I}(0:\infty)=\infty$ for any cut $(I,J)$. Thus, if a path from $J$ to $I$ exists in $\g_p$ (which automatically contains an arc $(j,i)$ with $j\in J,i\in I$) then a path (and in fact, an arc) from $I$ to $J$ exists as well.

In view of Remark~\ref{rem.unipos}, starting from some instant $k\ge k_0$ only persistent arcs exist, and, renumbering the agents, the matrix $W(k)$ becomes block-diagonal $W(k)=\diag(W_{11}(k),\ldots,W_{ss}(k))$, where $\{W_{jj}(k)\}_{k\ge k_0}$ are sequences of stochastic matrices corresponding to strongly connected graphs $\g_p^j$. Since the RAI decomposes
into several independent inequalities, it suffices to consider the case of a strongly connected persistent graph.

It remains to show that if $\g_p$ is strongly connected, then the RAI establishes consensus and~\eqref{eq.resid0} holds whenever the solution is bounded. This statement  is immediate from Corollary~\ref{cor.maximum}. Indeed, we know that the maximum $M(k)$ is non-increasing and hence converges to a limit $M_*=\lim_{k\to\infty}M(k)$. If $M_*=-\infty$, then all opinions reach consensus at $-\infty$. Otherwise, we have $x_i(k)\leq M(k)$ and thus
$\varlimsup_{k\to\infty}x_i(k)\leq M_*$. It suffices to show now that $\varliminf_{k\to\infty}x_i(k)\geq M_*$. Assume, on the contrary, that $\varliminf_{k\to\infty}x_i(k)< M_*-\ve$, where $\ve>0$. Then a sequence $k_m\to\infty$ exists such that $x_i(k_m)<M_*-\ve\leq M(k_m)-\ve$. In view of Corollary~\ref{cor.maximum}, there exists a sequence $k_m'>k_m$ such that $M(k_m')\leq M(k_m)-\varrho\ve$. Passing to the limit as $m\to\infty$, one arrives at $M_*\leq M_*-\varrho\ve$, which contradicts to the assumption $M_*>-\infty$. We have proved that $x_i(k)\xrightarrow[k\to\infty]M_*$ for every $i$, that is, $x(k)\xrightarrow[k\to\infty]{} M_*\ones_n$. Also, if $M_*>-\infty$, the residual can be represented as
\[
\begin{gathered}
0\leq\Delta(k)=W(k)x(k)-x(k+1)=\\=W(k)[x(k)-M_*\ones_n]-(x(k+1)-M_*)\xrightarrow[k\to\infty]{}0.
\end{gathered}
\]

The ``only if'' part in the last statement is immediate from Lemma~\ref{lem.persist1}.\epf

\subsection{Proof of Theorem~\ref{thm.static-ineq}}

To prove the ``if'' part of the first statement, it suffices to consider the case where $\g[W]$ is strongly connected (indeed, if the strong components of $\g[W]$ are isolated, then RAI~\eqref{eq.ineq} splits into several independent RAI).
If $\g[W]$ is strongly connected and aperiodic, then $W$ is a \emph{primitive} matrix~\citep{GantmacherVol2}, and hence $W^s$ has positive entries for sufficiently large $s\ge 0$, satisfying thus Assumptions~\ref{ass.aperiod} and~\ref{ass.unipos}. Since
$x(k+s)\leq W^sx(k)\quad\forall k\ge 0$, Theorem~\ref{thm.dynamic-ineq} (applied to $W(k)\equiv W^s$) entails that each subsequence $x(sm+j)$, $m=0,1,\ldots$ and $j=0,\ldots,s-1$, converges to a consensus vector
$x(sm+j)\xrightarrow[m\to\infty]{}c_j\ones_n$, $c_j\in [-\infty,\infty)$. By noticing that
$
x(sm+j+1)\leq Wx(sm+j),
$
one obtains that $c_0\ge c_1\ge\ldots c_{s-1}\ge c_0$, that is, $c_0=\ldots=c_{s-1}=c$ and the RAI establishes consensus.
The proof of~\eqref{eq.resid0} in the case where $c>-\infty$ is the same as in the time-varying case. This finishes the proof of the ``if'' part in the first and the last statements, as well as the two remaining statements of Theorem~\ref{thm.static-eq}.

To prove the converse statement, notice first that if the RAI~\eqref{eq.ineq} is convergent, the same holds for the RAI
\be\label{eq.RAI-s}
z(m+1)\leq W^sz(m),
\ee
where $s\ge 1$ is a fixed integer number. Indeed, if some solution of the latter RAI fails to have a limit, the same holds for the sequence
\[
x(k)=
\begin{cases}
z(m),\quad & k=ms\;\;\text{for some integer $m$}\\
Wx(k-1),\quad &\text{$s$ does not divide $k$},
\end{cases}
\]
which is a feasible solution to the RAI~\eqref{eq.ineq}. Also, any solution to the DeGroot's system~\eqref{eq.degroot} is a feasible solution to the RAI~\eqref{eq.ineq}. Hence, the RAI can be convergent only if all solutions of~\eqref{eq.degroot} converge, which means that each \emph{source} component of $\g[W]$ is an aperiodic graph~\citep{BulloBook-Online,ProTempo:2017-1}. For a source component of $\g[W]$ with the set of nodes $I$, $w_{ij}=0$ for any $i\in I$ and $j\not\in I$. Hence, the corresponding submatrix $W_I\dfb (w_{i,j})_{i,j\in I}$ is row-stochastic, irreducible and aperiodic (primitive). In particular, for $s$ being sufficiently large, the matrix $(W_I)^s$ (being a submatrix of $W^s$) has strictly positive entries.
Obviously, $I$ remains a source component in $\g[W^s]$; also, if there is an arc coming out of $I$ in $\g[W]$, that is, $w_{ji}>0$ for some $j\not\in I$ and $i\in I$, the same arc exists in the graph $\g[W^s]$ for large $s$.

In view of Remark~\ref{rem.aperiod}, Lemma~\ref{lem.persist1} is applicable to RAI~\eqref{eq.RAI-s} in spite of potential violation of Assumption~\ref{ass.aperiod}. Hence, all strong components of $\g[W^s]$ (in particular, all source components) are isolated for every sufficiently large $s$. Therefore, the \emph{source} components of $\g[W]$ are also isolated. As discussed in the proof of Lemma~\ref{lem.persist1}, this implies that in fact all components of $\g[W]$ are isolated (being simultaneously sources and sinks) and have to be aperiodic. If there is more than one strong component in the graph, then DeGroot's model~\eqref{eq.degroot} (and also the RAI) cannot provide consensus. This finishes the proof of ``only if'' parts in the first and the last statements.

\subsection{Proof of Theorem~\ref{thm.robust}}

The proof of Theorem~\ref{thm.robust}, except for its final statement, retraces the proof of Theorem~\ref{thm.dynamic-ineq}, where $M(k)$ is to be replaced by $\bar M(k)$ and
the proof of~\eqref{eq.resid0} for the case of a persistent graph $\g_p$ and a bounded solution has to be modified in the following way: if $x_i(k)\xrightarrow[]{}M_*>-\infty$, then
\[
\begin{gathered}
\Delta_i(k)=\sum_{j}w_{ij}(k)(x_j(k-d_{ij}(k))-M_*)-\\-(x_{i}(k+1)-M_*)\xrightarrow[k\to\infty]{}0.
\end{gathered}
\]
Lemma~\ref{lem.persist1} is applicable to the delayed case (Remark~\ref{rem.delay}).

To prove the final statement, concerned with the time-invariant situation, it suffices to consider the case where $\g[W]$ is strongly connected (the isolated strong components correspond to independent subsystems of inequalities). One can get rid of the delays by using a trick proposed by~\cite{Xiao2006}: consider the vector $y(k)\in\r^{d_*n}$ obtained by stacking the vectors $x(k),x(k-1),\ldots,x(k-d_*)$ one on top of another. Then it can be easily checked~\citep{Xiao2006} that
\[
y(k+1)\leq \Xi y(k),\quad \Xi=
\begin{bmatrix}
W_0 & W_1 & \ldots & W_{d_*-1} & W_{d_*}\\
 I_n & 0 & && 0\\
 0 & I_n && &0\\
 \vdots &&\ddots&&\vdots\\
 0 & 0 && I_n & 0
\end{bmatrix}
\]
Here $W_i$ are non-negative matrices such that $W=W_0+\ldots+W_{d_*}$; $(W_r)_{ij}=w_{ij}$ if and only if $d_{ij}=r$.
In particular, $\diag W_0=\diag W$ and, if $\diag W\ne 0$, the graph $\g[\Xi]$ has a self-arc and is aperiodic.
It remains to notice that each arc $(j,i)$ in the graph $\g[W]$ corresponds to a walk
$j\to (j+d_*)\to (j+2d_*)\to\ldots\to (j+d_{ij}d_*)\to i$ in the graph $\g[\Xi]$, therefore, the graph $\g[\Xi]$ is strongly connected. Applying Theorem~\ref{thm.static-ineq} to the matrix $\Xi$, one shows that RAI~\eqref{eq.ineq-d} establishes consensus. \epf

\subsection{Proof of Theorem~\ref{thm.new}}

The proof is based on the idea of ordering that was employed in analysis of continuous-time consensus algorithms over type-symmetric and cut-balanced weighted graphs~\citep{TsiTsi:13,MartinGirard:2013,MatvPro:2013} and later extended to the discrete-time case~\citep{Bolouki2015}. Note that the proofs from the aforementioned papers essentially use the boundedness of the solutions and their extension to a general feasible solution of RAI~\eqref{eq.ineq-d} is not straightforward. For this reason, we give an independent proof.

Let $\sigma(k)=(\sigma_1(k),\ldots,\sigma_n(k))$ denote the permutation of indices $1,\ldots,n$, sorting vector $x(k)$ in the ascending order
\[
y_1(k)\dfb x_{\sigma_1}(k)\leq y_2(k)\dfb x_{\sigma_2}(k)\leq\ldots\leq y_n(k)\dfb x_{\sigma_n}(k).
\]
Obviously, the sorted vector $y(k)$ obeys RAI
\be\label{eq.ineq-sorted}
y(k+1)\leq V(k)y(k),\quad v_{ij}(k)\dfb w_{\sigma_i(k+1)\sigma_j(k)}(k).
\ee
As shown in~\citep{Bolouki2015,XiaShiCao2019}, the stochastic matrix $V(k)$
also satisfies the uniform cut-balanced condition~\eqref{eq.cut-balance-dynam} (with a different constant $C$). Notice that this fact heavily relies on Assumption~\ref{ass.aperiod}.

\textbf{Step 1.} Using induction on $r=n,\ldots,1$, we will prove the following statements: \emph{1) the sequence $y_r(k)$ has a limit $\bar y_r\geq -\infty$ as $k\to\infty$; 2) if $\bar y_r>-\infty$, then}
\begin{gather}
\sum\nolimits_{k=0}^{\infty}v_{rs}(k)|y_r(k)-y_s(k)|<\infty\quad\forall s\in [1:n],\label{eq.l1}\\
\sum\nolimits_{k=0}^{\infty}v_{sp}(k)(y_r(k)-y_{r-1}(k))<\infty\,\forall s<r\,\forall p\geq r.\label{eq.l1+}
\end{gather}
(the condition~\eqref{eq.l1+} holds if $r>1$).


To prove the induction base $r=n$, recall that $y_n(k)=M(k)$ is a non-increasing function, which thus has a limit $\bar y_n$ as $k\to\infty$. Notice also that
\be\label{eq.aux9}
\begin{gathered}
y_n(k)-y_n(k+1)\geq \sum_{s=1}^{n-1} v_{ns}(k)(y_n(k)-y_s(k))\geq 0,
\end{gathered}
\ee
If $\bar y_n>-\infty$, the latter inequality, obviously, implies~\eqref{eq.l1} with $r=n$. To prove~\eqref{eq.l1+}, we apply the uniform cut-balance property to $I=[1:(n-1)], J=\{n\}$ and note that $y_n(k)-y_s(k)\geq y_{n}(k)-y_{n-1}(k)\,\forall s<n$, therefore
\[
\begin{gathered}
0\leq (y_{n}(k)-y_{n-1}(k))\sum_{s<n}v_{sn}(k)\overset{\eqref{eq.cut-balance-dynam}}{\leq}\\
\leq C(y_{n}(k)-y_{n-1}(k))\sum_{s<n}v_{ns}(k).
\end{gathered}
\]

To prove the induction step, assume that our statements have been already proved for $r=m+1,\ldots,n$. Our goal is to prove them for $r=m$. If $\bar y_{m+1}=-\infty$, then, obviously, $y_m(k)\xrightarrow[k\to\infty]{}-\infty$. Otherwise,~\eqref{eq.l1},\eqref{eq.l1+} hold for all $r>m$.
Using~\eqref{eq.l1+} for $s=m$ and any $p>m$, $r\in [(m+1):p]$,
\[
\begin{gathered}
v_{mp}(y_{p}-y_m)=v_{mp}(y_{m+1}-y_m)+v_{mp}(y_{m+2}-y_{m+1})+\\+\ldots+v_{mp}(y_{p}-y_{p-1})
\in\ell_1\quad\forall p>m
\end{gathered}
\]
(ss usual, we use $\ell_1$ to denote the set of all sequences $(a(k))$ such that $\sum_k|a(k)|<\infty$.)
Similar to~\eqref{eq.aux9}, we obtain that
\[
\begin{gathered}
y_m(k)-y_{m}(k+1)\geq \underbrace{\sum_{p>m}v_{mp}(k)(y_m(k)-y_p(k))}_{\in \ell_1}+\\
+\sum_{r\leq m}\underbrace{v_{mr}(k)(y_m(k)-y_r(k))}_{\geq 0},
\end{gathered}
\]
therefore, the exists the limit (possibly, infinite)
\[
\lim_{K\to\infty}(y_m(0)-y_m(K+1))=\sum_{k=0}^{\infty}(y_m(k)-y_m(k+1))\leq +\infty,
\]
in other words, $y_m(k)$ has a limit as $k\to\infty$. Furthermore, if this limit is finite, then also $v_{mr}(y_m-y_r)\in\ell_1$ for $r<m$, which proves~\eqref{eq.l1} for $r=m$.
Applying~\eqref{eq.l1} to all $s<m$ and $r=m,\ldots,n$ and noticing that $|y_r(k)-y_s(k)|\geq y_{m}(k)-y_{m-1}(k)$, one shows that
\[
(y_m-y_{m-1})\sum_{s<m}\sum_{r\geq m}v_{rs}(y_m-y_{m-1})\in\ell_1,
\]
and thus, applying~\eqref{eq.cut-balance-dynam} for $I=[1:(m-1)]$, $J=[m:n]$, we also have $v_{sp}(y_m-y_{m-1})\in\ell_1$
if $s<m\leq p$. This proves~\eqref{eq.l1+} for $r=m$ and finishes the induction step.

\textbf{Step 2.} Notice that the convergence of the sorted vectors $y(k)$ in general \emph{does not} imply the convergence of the original vector $x(k)$. A trivial counterexample is the 2-periodic sequence $x(0)=(1,0)^{\top},x(1)=(0,1)^{\top},x(k)=x(k-2)\,\forall k\geq 2$. Obviously, $y(k)\equiv x(0)$, whereas $x(k)$ fails to converge. In order to prove convergence of $x(k)$, we have to use the aperiodicity property (Assumption~\ref{ass.aperiod}).

Suppose first that $\bar y_1>-\infty$, that is, the solution is bounded. Then,~\eqref{eq.l1} holds for every $r\geq 1$, entailing that $\sum_{k=0}^{\infty}|y_r(k+1)-y_r(k)|<\infty$ and thus
\[
\begin{gathered}
\sum_{k=0}^{\infty}\sum_{i,j=1}^nw_{ij}(k)|x_i(k+1)-x_j(k)|\leq\\
\leq\sum_{k=0}^{\infty}\sum_{s,r=1}^nv_{rs}(k)|y_r(k+1)-y_s(k)|<\infty
\end{gathered}
\]
(recall that $v_{rs}(k)=w_{ij}(k)$ if and only if $r=\sigma_i(k+1)$ and $s=\sigma_i(k)$, for this reason we have $x_i(k+1)=y_r(k+1)$ and $x_j(k)=y_s(k)$). Since $w_{ii}(k)$ are uniformly positive, we have $\sum_{k=0}^{\infty}|x_i(k+1)-x_i(k)|<\infty$ (therefore, each $x_i(k)$ converges to a finite limit $\bar x_i$) and  $w_{ij}(x_i-x_j)\in\ell_1$ for all $i,j$, which implies that $\bar x_i=\bar x_j$ if $(j,i)\in\mathcal{E}_p$, that is,
consensus in each strongly connected component of $\g_p$ is established. To prove~\eqref{eq.resid0}, notice that
\be\label{eq.delta-l1}
\Delta_i(k)=\sum_jw_{ij}(k)(x_j(k)-x_i(k))-(x_i(k+1)-x_i(k)),
\ee
which sequence belongs to $\ell_1$.

The case of unbounded solution is more sophisticated. Assume that $\bar y_1=\ldots=\bar y_m=-\infty<\bar y_{m+1}$.
Notice first that if a sequence $k_m\to\infty$ such that the limit $\xi=\lim_{m\to\infty}x_i(k_m)$ exists, then $\xi$ is one of the values $\bar y_1,\ldots,\bar y_n$. Indeed, for every $m$ we have $x_i(k_m)=y_{j_m}(k_m)$ with for some $j_m$.
Passing to the subsequence, we may assume, without loss of generality, that $j_m\equiv j$, and thus $\xi=\bar y_j$.
For the same reason, for every $\bar y_j$ there exists at least one sequence $x_i(k_m)\xrightarrow[m\to\infty]{}\bar y_j$: indeed, $y_j(k)=x_{\sigma_j(k)}(k)$ for some reason, and we can find a subsequence $k_m$ such that $\sigma_j(k_m)\equiv i$ for all $m$.

Let $J$ stand for the set of indices $i$ such that
\[
\varliminf\limits_{k\to\infty}x_i(k)=-\infty,
\]
and $I=J^c$. It can be easily seen that $x_j(k)\to-\infty$ for every $j\in I$ and $\varliminf_{k\to\infty} x_i(k)\geq \bar y_{m+1}$ for all $i\in I$. Indeed, choose two constants $b<\bar y_{m+1}$ and $a<b$ in such a way that $\eta a+(1-\eta)\bar y_n<b$. For large $k$, we have $y_m(k)<a<b<y_{m+1}(k)$, therefore, no component $x_i(k)$ can belong to $[a,b]$.
Hence, if $x_j(k)<a$ for $k$ sufficiently large, RAI~\eqref{eq.ineq} and Assumption~\ref{ass.aperiod} imply that
\[
x_j(k+1)<\eta a+(1-\eta)y_n(k)<b\Longrightarrow x_j(k+1)<a.
\]
that is, the opinion cannot leave the interval $(-\infty,a)$. Therefore, for every $j\in J$ one has $\varlimsup_{k\to\infty} x_j(k)\leq a$, which is possible only if $x_j(k)\to -\infty$. For $i\in I$, we have $\varliminf_{k\to\infty} x_i(k)\geq a$, which is possible only if $\varliminf_{k\to\infty} x_i(k)\geq y_{m+1}$.
Obviously, $J=\{\sigma_1(k),\ldots,\sigma_m(k)\}$ for $k$ being sufficiently large, whereas $I=\{\sigma_{m+1}(k),\ldots,\sigma_n(k)\}$. Now, we can repeat the argument we used in the bounded solution case with a minor modification. Applying~\eqref{eq.l1} for every $r>m$, one has
\[
\begin{gathered}
\sum_{i\in I}\sum_{j=1}^nw_{ij}(k)|x_i(k+1)-x_j(k)|=\\
=\sum_{r>m}\sum_{s=1}^nv_{rs}(k)|y_r(k+1)-y_s(k)|,
\end{gathered}
\]
for $k$ being sufficiently large, and therefore the latter sequence belongs to $\ell_1$. Since $w_{ii}(k)\geq\eta$, we have $\sum_{k=0}^{\infty}|x_i(k+1)-x_i(k)|<\infty$ for $i\in I$ (therefore, $x_i(k)$ converges to a finite limit $\bar x_i$ for all $i\in I$) and $\sum_{k=0}^{\infty}w_{ij}(k)|x_j(k)-x_i(k)|<\infty$ for every $i\in I,j\in [1:n]$.
Using~\eqref{eq.delta-l1}, one proves that $\Delta_i\in \ell_1$ for $i\in I$. Obviously, if $i\in I$ and $j\in J$, then $(j,i)\not\in\mathcal{E}_p$, that is, there are no arcs between $I$ and $J$ in $\mathcal{E}_p$ (recall that $\g_p$ automatically has isolated strongly connected components in view of the cut-balance property). We know that $\bar x_j=-\infty$ for $j\in J$, if $i,j\in I$ and $(j,i)\in\mathcal{E}_p$, then $\bar x_i=\bar x_j$, so that consensus in each strongly connected component of $\g_p$ is established.\epf

\begin{remark}
Using techniques from~\cite{XiaShiCao2019}, the consensus criterion from Theorem~\ref{thm.new} can be easily extended to a more general situation of non-instantaneous cut-balance~\eqref{eq.cut-balance-dynam1}. Namely, if~\eqref{eq.cut-balance-dynam1} holds, Assumption~\ref{ass.aperiod} is valid and $\g_p$ is strongly connected, then RAI~\eqref{eq.ineq} establishes consensus. Indeed, matrices $B(k)=W(k+L-1)\ldots W(k+1)W(k)$ satisfy the uniform cut-balanced condition~\eqref{eq.cut-balance-dynam} and Assumption~\ref{ass.aperiod}~\citep{XiaShiCao2019}. Since
\[
x(j+(k+1)L)\leq B(j+kL)x(j+kL),
\]
for each $j=0,\ldots,L-1$, the sequence $x(j+kL)$ converges to consensus vector $c_j\ones_n$ as $k\to\infty$. Similar to the proof of Theorem~\ref{thm.static-eq}, one proves that $c_0=c_1=\ldots=c_{L-1}$.
\end{remark} 

\section{Conclusions and Perspectives}\label{sec.conclus}
The recurrent averaging inequalities (RAIs) introduced in~\eqref{eq.ineq} can be considered as a relaxed form of the conventional dynamics of
iterative averaging~\eqref{eq.degroot}.  
While in the standard iterative averaging  each agent updates its opinion with a weighted average of its own and the others' opinions, in the RAI each agent is allowed to choose \emph{any} opinion which does not exceed that linear combination.
Such a constraint seems only loosely restrictive, however, under some connectivity assumptions, all feasible solutions converge and even reach consensus. Similarly to the classical iterative averaging, consensus is robust to bounded communication delays.
Of particular interest are  the bounded solutions to the RAI for which the residual vector (the difference between the right-hand and the left-hand sides) vanishes asymptotically.

The systematic study of RAIs appears to be relevant since they naturally arise, implicitly or explicitly, in a number of distributed algorithms and multi-agent models based on iterative averaging.
Using the general results on consensus in RAIs, we  derive a number of known and new results in a unified way, namely,
a stability criterion for substochastic matrices and its novel extension to time-delay systems, the convergence of opinions in the Hegselmann-Krause model with ``truth-seekers'' and the discrete-time model of opinion polarization over a signed network, and the convergence of algorithms for finding a common fixed point for a family of paracontractions.
The latter algorithms are quite relevant and can be used, in particular, for solving in a distributed way large-scale systems of equations and strongly convex optimization problems~\citep{FullmerMorse2018,Wang2019_ARC}.

The results in this paper can potentially be extended to RAIs with random matrices $W(k)$, and to RAIs obtained from \emph{nonlinear} procedures of iterative averaging~\citep{Moro:05,Bliman:06,Fang:08,ProMatv:15}. However, the corresponding results require additional mathematical tools and are thus beyond the scope of the present work.

Although the theory of RAIs allows a unified analysis of  many multi-agent algorithms and models that are based on iterative averaging, some consensus-based algorithms  still require an ad-hoc  theory.
One class of  such algorithms is constituted by distributed optimization procedures employed, in particular, in machine learning, multi-sensor tracking, control of cyber-physical systems and smart power grids, see, e.g.,~\citep{Nedic:10,Bajovic:2017,Tatarenko:2017,Mokhtari:16,Molzahn:2017,EROFEEVA:2018,Notarstefano-Survey2019,Scaman:2019,YANG2019_ARC} and the references therein. Another example is given by proximal dynamics in multi-agent games~\citep{Grammatico:2018,CenedeseKawano:2018}, which are similar in spirit to the algorithms in Section~5.4, but the corresponding proximal maps do not have a common fixed point.
We believe, however, that extensions of RAIs theory towards such algorithms and models are possible and should be
the subject of  future research.

\bibliographystyle{elsarticle-harv}
\bibliography{consensus,social}

\end{document}